\newcommand{\be}{\begin{equation}}
\newcommand{\ee}{\end{equation}}
\newcommand{\bea}{\begin{eqnarray}}
\newcommand{\eea}{\end{eqnarray}}
\newcommand{\nn}{\nonumber}
\newcommand{\Appendix}[1]%
    {\renewcommand{\thesection}{Appendix~\Alph{section}:}%
         \section{#1}}%
\catcode`@=11
\long\def\@makecaption#1#2{
   \vskip 10pt
   \setbox\@tempboxa\hbox{{\small\bf #1.} \ {\small #2}}
   \ifdim \wd\@tempboxa >\hsize       % IF longer than one line:
   {\small\bf #1.} \ {\small #2}\par  % THEN set as ordinary paragraph.
   \else                              %   ELSE  center.
        \hbox to\hsize{\hfil\box\@tempboxa\hfil}
   \fi}
\catcode`@=12

\catcode`@=11
\def\secteqno{\@addtoreset{equation}{section}%
\def\theequation{\thesection.\arabic{equation}}}
\def\endsecteqno{\def\theequation{\@ifundefined{chapter}%
{\arabic{equation}}{\thechapter.\arabic{equation}}}}
\newcounter{subequation}
\def\thesubequation{\alph{subequation}}
\def\sneqnarray{\stepcounter{equation}\let\@currentlabel=\theequation
\setcounter{subequation}{1}
\def\@eqnnum{{\rm (\theequation\thesubequation)}}
\global\@eqcnt\z@\tabskip\@centering\let\\=\@eqncr\let\@@eqncr=\@@sneqncr
$$\halign to \displaywidth\bgroup\@eqnsel\hskip\@centering
 $\displaystyle\tabskip\z@{##}$&\global\@eqcnt\@ne
 \hskip 2\arraycolsep \hfil${##}$\hfil
 &\global\@eqcnt\tw@ \hskip 2\arraycolsep
$\displaystyle\tabskip\z@{##}$\hfil
tabskip\@centering&\llap{##}\tabskip\z@\cr}
\def\endsneqnarray{\@@sneqncr\egroup $$\global\@ignoretrue}
\def\@@sneqncr{\let\@tempa\relax
   \ifcase\@eqcnt \def\@tempa{& & &}\or \def\@tempa{& &}
   \else \def\@tempa{&}\fi
     \@tempa \if@eqnsw\@eqnnum\stepcounter{subequation}\fi
     \global\@eqnswtrue\global\@eqcnt\z@\cr}
\def\nobiblabels{\def\@lbibitem[##1]##2{\@bibitem{##2}}}
\catcode`@=12

\def\beq{\begin{equation}}
\def\eeq{\end{equation}}
\def\bea{\begin{eqnarray}}
\def\eea{\end{eqnarray}}
\def\nn{\nonumber}

\def\la{\lambda} \def\lap{\lambda^{\prime}}  \def\de{\delta}  \def\dag{\dagger}
   \def\Oc{{\rm O}} \def\S{{\rm S}}
\def\bnabla{{\bm \nabla}}
\def\bsigma{{\bm \sigma}}

\documentclass[aps,10pt,prd,showpacs,amsmath,amssymb,preprintnumbers,superscriptaddress,nofootinbib,showkeys]{revtex4-1}

\usepackage[german,english]{babel}
\usepackage{hyperref}
\usepackage{amssymb}
\usepackage{amsmath}
\usepackage{bm}% bold math   
\usepackage{braket}
\usepackage{slashed}
\usepackage{multirow}
\usepackage{epsfig}
\usepackage{epstopdf}
\usepackage{bbm}
\usepackage{bbold}
\usepackage{diagbox}
\usepackage{soul}%remove eventually

\begin{document}

\title{Exotic to standard bottomonium transitions}

\preprint{JLAB-THY-21-3349}

\author{Jaume Tarr\'us Castell\`a}
\email{jtarrus@ifae.es}
\affiliation{Grup de F\'\i sica Te\`orica, Dept. F\'\i sica and IFAE-BIST, Universitat Aut\`onoma de Barcelona,\\ 
E-08193 Bellaterra (Barcelona), Catalonia, Spain}

\author{Emilie Passemar}
\email{epassema@indiana.edu}
\affiliation{Department of Physics, Indiana University, Bloomington, Indiana 47408, USA}
\affiliation{Center for Exploration of Energy and Matter,
Indiana University, Bloomington, Indiana 47408, USA}
\affiliation{Theory Center, Thomas Jefferson National
Accelerator Facility, Newport News, Virginia 23606, USA}

\date{\today}

\begin{abstract}
We study the transition widths of $\Upsilon(10753)$ and $\Upsilon(11020)$ into standard bottomonium under the hypothesis that they correspond to the two lowest laying $1^{--}$ hybrid bottomonium states. We employ weakly coupled potential NRQCD an effective field theory incorporating the heavy-quark and multipole expansions. We consider the transitions generated by the leading order and next-to-leading order singlet-octet operators. In the multipole expansion the heavy-quark matrix elements factorize from the production of light-quark mesons by gluonic operators. For the leading order operator we compute the widths with a single $\pi^0$, $\eta$ or $\eta'$ in the final state and for the next-to-leading operator for $\pi^+\pi^-$ or $K^+K^-$. The hadronization of the gluonic operators is obtained, in the first case, from the axial anomaly and a standard $\pi^0-\eta-\eta'$ mixing scheme and, in the second case, we employ a coupled-channel dispersive representation matched to chiral perturbation theory for both the $S$- and $D$-wave pieces of the gluonic operator. We compare with experimental values and semi-inclusive widths. Our results strongly suggest that $\Upsilon(11020)$ is indeed a hybrid bottomonium state. 
\end{abstract}

\maketitle

\section{Introduction}

Hadrons have been traditionally classified according to their number of valence quarks. In the naive quark model, three quark states are called baryons while quark-antiquark states are called mesons. Even from the inception of the quark model, it was noted that more complex states, for example containing four or five quarks, were possible~\cite{GellMann:1964nj}. Another possibility, unique to QCD, is the participation of gluons as valence degrees of freedom. The so-called hybrid states are the ones combining both quark and gluonic valence degrees of freedom. Such nonconventional states, often referred as exotics, were absent from the experimental observations up to 2003 when the Belle experiment discovered the $X(3872)$~\cite{Choi:2003ue}. This opened a period, up until present times, with the continuous discovery of new exotic states, particularly in the double heavy-quark sector.

Heavy quarks in hadrons are nonrelativistic and therefore their number is well defined. Heavy-quark-antiquark states, called quarkonium, are nonrelativistic bound states with quantum numbers akin to the hydrogen atom and are well understood. Therefore, when new states appeared in the charmonium and bottomonium spectrum that did not fit the standard quarkonium expectations, these were clear candidates to exotic states. In some cases the new states had explicitly exotic quantum numbers, such as the charged exotic quarkonium states which must include two heavy and two light quarks.

Several proposals have been made concerning the structure of the exotic quarkonium states: heavy hybrids, compact tetraquarks, hadro-quarkonium and heavy meson molecules. Moreover, for each of these pictures several theoretical approaches can be found in the literature. Some common predictions from these pictures and approaches are the spectrum and the composition of heavy-quark spin symmetry multiplets. However, often several interpretations are consistent with the observed spectrum and not enough quantum numbers of exotic quarkonium are accessible experimentally to be able to check heavy-quark spin symmetry multiplet predictions. Another avenue to understand the structure of exotic quarkonium is the study of their decays, in particular transitions into standard quarkonium states with one or two light-quark mesons in the final state, since many of the known exotic quarkonium states have been discovered through these decay channels.

The objective of this paper is to study a set of exotic to standard quarkonium transitions in a nonrelativistic effective field theory (EFT) approach. Since the heavy quarks in exotic quarkonium are nonrelativistic, the natural starting point for their study is NRQCD~\cite{Caswell:1985ui,Bodwin:1994jh} at leading order, that is in the static limit. In this limit the spectrum is composed of the so-called static energies, which depend on the quantum numbers of the light quarks and gluon degrees of freedom, the heavy-quark antiquark distance, and the representation of the cylindrical symmetry group $D_{\infty h}$.\footnote{See, for example, Appendix~A of Ref.~\cite{Berwein:2015vca} for a detailed description of the $D_{\infty h}$ group.} The static energies are nonperturbative quantities that should be computed in lattice QCD. So far only the spectrum in the quenched approximation is known~\cite{Juge:2002br,Capitani:2018rox}. Going beyond the static limit, heavy-quark-antiquark bound states are formed around the minima of the static energies. These states correspond to the exotic quarkonium and in the case of the static energies from Refs.~\cite{Juge:2002br,Capitani:2018rox} to the hybrid quarkonium picture. Since the heavy-quark-antiquark binding energy is much smaller than the energy scale that characterizes the static energies, $\Lambda_{\rm QCD}$, one can write an EFT describing hybrid quarkonium~\cite{Berwein:2015vca,Brambilla:2017uyf,Oncala:2017hop,Brambilla:2018pyn,Brambilla:2019jfi}, which at leading order coincides with the Born-Oppenheimer approximation for heavy hybrids~\cite{Griffiths:1983ah,Juge:1999ie,Guo:2008yz,Braaten:2014qka,Capitani:2018rox,Mueller:2019mkh}. This kind of Born-Oppenheimer EFT has been generalized to any light-quark and gluon states in Ref.~\cite{Soto:2020xpm} and also to double heavy-quark states such as double heavy baryons~\cite{Soto:2020pfa}.

Although the precise spectrum of heavy-quark-antiquark static energies with dynamical light quarks is not known, we do have pieces of information from lattice studies to infer a general picture. In Ref.~\cite{Bali:2000vr} the ground and first excited static energies were obtained both in the quenched and unquenched computations with no significant differences encountered. We expect this to hold for the rest of the static energies computed in Refs.~\cite{Juge:2002br,Capitani:2018rox}. Nevertheless, with dynamical light quarks new static states appear, most importantly heavy-meson pairs. In fact, many exotic quarkonium states have been interpreted as heavy-meson shallow bound states, see Ref.~\cite{Guo:2017jvc} for a review on the topic. The effect of these thresholds on the hybrid quarkonium states can be assessed from the string breaking studies in lattice QCD~\cite{Bali:2005fu,Bulava:2019iut} which suggest that threshold effects are only noticeable in a tiny energy band around the threshold of a few tens of MeV. The emergent picture is that the hybrid states, as described in the previous paragraph, are a good approximation of a more general isospin $I=0$ exotic quarkonium states.

To study the exotic to standard quarkonium transitions we will employ the multipole expansion. For heavy-quark-antiquark systems the EFT that incorporates the multipole expansion is weakly coupled potential NRQCD (pNRQCD)~\cite{Pineda:1997bj,Brambilla:1999xf}. Unfortunately, the multipole expansion is a poor expansion for hybrid charmonium~\cite{Berwein:2015vca} and even for standard charmonium states beyond the ground state~\cite{Peset:2018jkf}. For this reason, in this paper we will restrict ourselves to the bottomonium sector. In this sector we encounter the following exotic states: three neutral $1^{--}$ states: $\Upsilon(10753)$, $\Upsilon(10860)$ and $\Upsilon(11020)$\cite{Santel:2015qga,Abdesselam:2015zza,Abdesselam:2019gth}; and two charged $1^+$ ones: $Z_b(10610)$ and $Z_b(10650)$~\cite{Belle:2011aa}. Both the charged ones and $\Upsilon(10860)$ lay very close to B meson pair thresholds and therefore are very likely to be molecular states. Nevertheless the $\Upsilon(10860)$ could have a small $\Upsilon(5S)$ component. Our interest will be in the states $\Upsilon(10753)$ and $\Upsilon(11020)$ which we will identify as the two lowest $1^{--}$ hybrid bottomonium states. The predictions for these states from Ref.~\cite{Berwein:2015vca} are $10.79$ and $10.98$~GeV. For the first state the difference is of $40$~MeV, which is significant, however this is off only a handful of data points from Ref.~\cite{Abdesselam:2019gth}, and it is possible that future data might change the mass of this state by an amount of this order. For the second state the difference is of $20$~MeV which is well within the uncertainties of the theoretical prediction.

We will investigate the transitions of $\Upsilon(10753)$ and $\Upsilon(11020)$ into standard bottomonium in weakly coupled pNRQCD in a similar approach to the one used in Ref.~\cite{Pineda:2019mhw} to study transitions in standard quarkonium; that is, we will assume the following hierarchy of scales $m_Q\gg m_Q v\gg \Lambda_{\rm QCD}$ is fulfilled, with $m_Q$ the heavy-quark mass and $v$ the relative heavy-quark-antiquark velocity. We will provide predictions for the transition widths when the final state includes a single $\pi^0$, $\eta$ or $\eta'$ or a pair of $\pi^+\pi^-$ or $K^+K^-$. The matrix elements for the single meson production are obtained using the $U(1)_A$ anomaly and the Feldmann-Kroll-Stech (FKS) $\pi^0$-$\eta$-$\eta'$ mixing scheme~\cite{Feldmann:1998vh,Kroll:2005sd}. Unlike Ref.~\cite{Pineda:2019mhw}, the two pion and two kaon matrix elements are not obtained through a chiral representation, since the large mass difference between the initial exotic state and the final standard quarkonium makes that unfeasible. Instead we build a dispersive representation of the relevant gluonic matrix elements that takes into account the pion and kaon scattering as well as their coupling. This is analogous to the approach in Refs.~\cite{Donoghue:1990xh,Moussallam:1999aq,Celis:2013xja} for the $S$-wave part of the matrix element but is also applied for the first time to the $D$-wave piece.

The paper is organized as follows. In Sec.~\ref{s1} we introduce the pNRQCD Lagrangian and define the standard and hybrid quarkonium states. In Secs.~\ref{s2} and \ref{s3} we study the leading order (LO) and next-to-leading order (NLO) transitions, respectively, and provide numerical predictions for the widths of $\Upsilon(10753)$ and $\Upsilon(11020)$ into a set of specific light-quark final states. We discuss several ratios in Sec.~\ref{s4} for which some uncertainties cancel out. In Sec.~\ref{s:siw} we compute the semi-inclusive width for the transitions that allow it and discuss the results in relation to the ones in Secs.~\ref{s2} and \ref{s3}. We give our conclusions in Sec.~\ref{s5}. In Appendix~\ref{Ap:krollm} we review the computation of the $\pi^0,\,\eta$ and $\eta'$ production matrix elements with the $U(1)_A$ anomaly and the FKS mixing scheme. In Appendix~\ref{Ap:om} we build the dispersive representation for the two-pion and two-kaon production form factors. Finally, in Appendix~\ref{Ap:mvw}, we collect the definitions of the Mandelstam variables and several formulas employed in the evaluation of widths from the transition amplitudes.

\section{Standard and Hybrid quarkonium in \texorpdfstring{\lowercase{p}}{p}NRQCD}\label{s1}

\subsection{\texorpdfstring{\lowercase{p}}{p}NRQCD Lagrangian}

The pNRQCD Lagrangian at LO in $1/m_Q$, where $m_Q$ is the heavy-quark mass, except for the kinetic term and at LO in the multipole expansion reads
\begin{align}
L^{\rm LO}_{\rm pNRQCD} =& \int d^3R\Bigg\{\int d^3r \,{\rm Tr}\left[\S^{\dag}\left(i\partial_0-h^{(0)}_s\right)\S+ \Oc^{\dagger}\left(iD_0-h^{(0)}_o\right)\Oc\right] %\nn \\
%&
-\frac{1}{4} G_{\mu \nu}^a G^{\mu \nu\,a} + \sum^{n_f}_{i=1}\bar{q}_i (i\slashed{D} -m_i)q_i
\Biggr\}\,.
\label{pnrqcd1}
\end{align}
 $\S$ and $\Oc$ are the quark singlet and octet fields, respectively, normalized with respect to color as $\S=S \bm{1}_c/\sqrt{N_c}$ and $\Oc=O^a T^a/\sqrt{T_F}$. The dependence in $t$, the relative coordinates $\bm{r}$, and the center of mass coordinates $\bm{R}$ of the heavy quarks of the singlet and octet fields is left implicit. The trace should be understood as a double trace in color and spin. The singlet and octet fields are organized in $SU(2)$ spin multiplets. For instance, $S=({\bf S}\cdot {\bsigma}+S_{\eta}\mathbb{1}_2)/\sqrt{2}$. All the fields of the light degrees of freedom in Eq.~\eqref{pnrqcd1} are evaluated at $\bm{R}$ and~$t$; in particular, $G^{\mu \nu\,a}\equiv G^{\mu\nu\,a}(\bm{R},\,t)$, $q_i\equiv q_i(\bm{R},\,t)$, and  $iD_0 O\equiv i \partial_0O-g\left[A_0(\bm{R},\,t),O\right]$. The singlet and octet Hamiltonian densities read as
\begin{align}
h^{(0)}_s=&-\frac{\bnabla^2_r}{m_Q}+V^{(0)}_s(r)\,,\\
h^{(0)}_o=&-\frac{\bnabla^2_r}{m_Q}+V^{(0)}_o(r)\,,
\end{align}
where $V^{(0)}_s(r)$ and $V^{(0)}_o(r)$ are computed in perturbation theory. Note that we have spin symmetry. 

At NLO in the multipole expansion or in $1/m_Q$ we have the following operators that produce transitions between singlet and octet fields
\begin{align}
L^{\rm NLO}_{\rm pNRQCD} =\int d^3R d^3r\Bigg\{g{\rm Tr}\left[\S^{\dag}\bm{r}\cdot\bm{E}\,\Oc+\Oc^{\dag}\bm{r}\cdot\bm{E}\,\S\right]+\frac{gc_F}{m_Q}{\rm Tr}\left[\S^{\dag}(\bm{S}_1-\bm{S}_2)\cdot\bm{B}\,\Oc+\Oc^{\dag}(\bm{S}_1-\bm{S}_2)\cdot\bm{B}\,\S\right]\Bigg\}\,.
\label{pnrqcd2}
\end{align}  
The spin vectors $\bm{S}_1$ and $\bm{S}_2$ correspond to the heavy-quark and heavy-antiquark respectively. The chromoelectric and chromomagnetic fields are defined as $\bm{E}^i=G^{i0}$ and $\bm{B}^i=-\epsilon_{ijk}G^{jk}/2$ with $\epsilon_{123}=1$.

\subsection{Standard quarkonium states}

Now, let us define the standard quarkonium states. In the static limit these are simply
\begin{align}
|{\bf R}, {\bf r};\Sigma_g^+\rangle =S^{\dagger}\left(\bm{R},\bm{r}\right)|0\rangle\,.
\end{align}
The full static potential corresponds to the static energy in the $\Sigma_g^+$ representation 
\begin{align}
V^{(0)}_{\Sigma_g^+}(r)&=\lim_{t \rightarrow \infty}\frac{i}{t}\ln \langle{\bf R}, {\bf r};\Sigma_g^+; t/2 |{\bf R}, {\bf r};\Sigma_g^+;-t/2\rangle=V^{(0)}_s+b_{\Sigma_g^+} r^2+\cdots=E^{(0)}_{\Sigma_g^+}(r)\,.\label{ssp}
\end{align}
We are going to use a fit to the full static energy as the static potential instead of a the multipole expanded expression in the middle equality in Eq.~\eqref{ssp} in order to increase accuracy. The lattice data used for $\Sigma_g^+$ and the fitted potential can be found in Fig.~\ref{potentials}. The static eigenstates can be used as a basis for a general quarkonium state
\begin{align}
&|S_m\rangle=\int d^3\bm{r}d^3\bm{R}\,\phi^{(m)}(\bm{R}, \bm{r})|\bm{R}, \bm{r};\Sigma_g^+\rangle\,,
\end{align}
then, we can use quantum mechanical perturbation theory to incorporate the kinetic term and obtain the Shr\"odinger equation for the standard quarkonium states.
\begin{align}
&\left(-\frac{\bnabla^2_r}{m_Q}+V^{(0)}_{\Sigma_g^+}(r)\right)\phi^{(m)}(\bm{r})={\cal E}_m \phi^{(m)}(\bm{r})\,,
\end{align}
where we have used the short-hand notation for the wave function
\begin{align}
\phi^{(m)}(\bm{r})\equiv\phi^{m j l s }(\bm{r})=\phi_m(r)\Phi^0_{^{2s+1}l_j}(\theta,\phi)\,,
\end{align}
with $m$ the principal quantum number, $l(l+1)$ the eigenvalue heavy-quark pair angular momentum $\bm{L}^2_{\bar{Q}Q}$, $s(s+1)$ the heavy-quark pair spin $\bm{S}^2=\left(\bm{S}_1+\bm{S}_2\right)^2$ eigenvalue and $j(j+1)$ the eigenvalue of the total angular momentum $\bm{J}^2=(\bm{L}_{\bar{Q}Q}+\bm{S})^2$. For brevity, we will use $(m)$ to denote the whole set of quantum numbers.

\subsection{Hybrid quarkonium states}

The hybrid states are more complicated due to the presence of nontrivial gluonic degrees of freedom. At LO in the heavy-quark mass and multipole expansions the gluonic excitations are characterized by local operators belonging to irreducible representations of the $O(3)\otimes$C group which are called gluelump operators. We write these operators as $G_{k}^{ia}(\bm{R})$ where $a$ is the color index, $k$ labels the gluelump $J^{PC}$ and $i$ labels its vector components. Let $H^{(0)}$ be the LO Hamiltonian density in the multipole and heavy-quark mass expansions corresponding to the Lagrangian in Eq.~\eqref{pnrqcd1}
\begin{align}
H^{(0)}
&=\int d^3{\bf R}\int d^3{\bf r} \, {\rm Tr}\left[\S^{\dag}V^{(0)}_s\S+ \Oc^{\dagger}V^{(0)}_o\Oc\right]
+ \int d^3{\bf R} \left( \frac{1}{2} \left(\bm{E}^a\cdot\bm{E}^a+\bm{B}^a\cdot\bm{B}^a\right)
- \sum^{n_i}_{i=1}\bar{q}_i \,[ i  \bm{D}\cdot{\bm \gamma} -m_i]\, q_i \right)\,.
\label{h-total}
\end{align}
The gluelump operators are the eigenstates $H^{(0)}$ in the presence of a local heavy-quark-antiquark octet source:
\begin{align}
H^{(0)} G^{ia}_{k}(\bm{R})O^{a \dagger}({\bf R},{\bf r}) |0\rangle = (\Lambda_{k}+V_o^{(0)}) 
G^{ia}_{k}(\bm{R})O^{a \dagger}({\bf R},{\bf r})|0\rangle\,.
\end{align}
The gluelump operators are normalized as
\begin{align}
\langle 0|G_k^{ia\,\dagger}(\bm{R'})\,O^{a}(\bm{R'},{\bf r}')O^{b\,\dagger}(\bm{R},{\bf r})G_{k^{\prime}}^{jb}(\bm{R})|0\rangle=\de^{ij}\de_{kk^{\prime}}\delta({\bf R'}-{\bf R})\delta({\bf r'}-{\bf r})
\,.\label{glmpn}
\end{align}
For simplicity from now on we will only consider the lowest laying gluelump operator with $k=1^{+-}$~\cite{Foster:1998wu}, $G^{ia}_{B}\equiv G^{ia}_{1^{+-}}$, which is the one associated to the lowest laying hybrid states. Once we go beyond the short-distance limit the symmetry of the system is reduced to the cylindrical symmetry group $D_{\infty h}$. One can construct gluonic operators in irreducible representations
of this group by contracting the gluelump operator with appropriate projection vectors, hence the hybrid static states can be written as
\begin{align}
|{\bf R}, {\bf r};\lambda\rangle =\bm{\hat{r}}_{\lambda}\cdot\bm{G}_{B}^{a}(\bm{R})O^{a\,\dagger}\left(\bm{R},\bm{r}\right)|0\rangle\,,
\label{eigen1}
\end{align}
with the projector vectors 
\begin{align}
&\bm{\hat{r}}_0= \bm{\hat{r}}\,,\label{pr10}\\
&\bm{\hat{r}}_{\pm1}=\mp\left(\bm{\hat{\theta}}\pm i\bm{\hat{\phi}}\right)/\sqrt{2}\,,\label{pr11}
\end{align}
where $\bm{\hat{\theta}}=\left(\cos\theta\cos\varphi,\cos\theta\sin\varphi,-\sin\theta\right)^T$ and $\bm{\hat{\varphi}}=\left(-\sin\varphi,\cos\varphi,0\right)^T$ are the usual local unit vectors in a spherical coordinate system. The quantum number $\lambda=0,\pm 1$ is the eigenvalue of the projection of the gluelump spin into the heavy-quark-antiquark axis.

\begin{figure}[ht!]
   \centerline{\includegraphics[width=.5\textwidth]{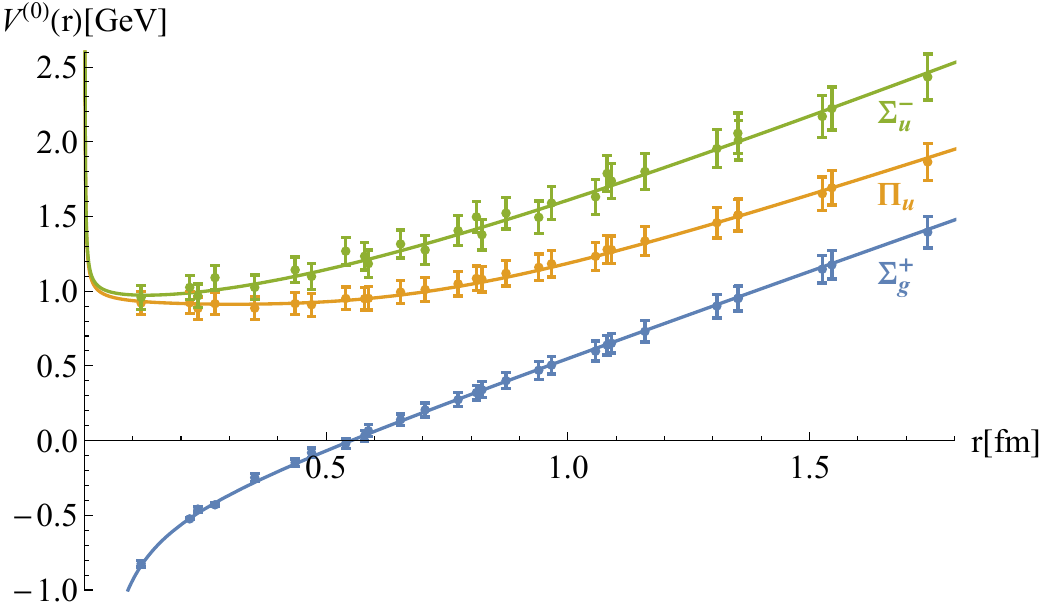}}
	\caption{Lattice data on the heavy-quark-antiquark static energies $\Sigma^+_g$, $\Pi_u$ and $\Sigma^-_u$ in the quenched approximation of Ref.~\cite{Juge:2002br}. The curves correspond to the fitted potentials used in the numerical solution of the Schr\"odinger equations to find the standard and hybrid states wave functions. Different energy offsets are added to the $\Sigma^+_g$ data and to $\Pi_u$ and $\Sigma^-_u$ data. The first is chosen so the ground state mass matches the experimental $\Upsilon(1S)$ mass and the second is chosen so the lattice data matches the short distance expansion in Eq.~\eqref{Vkappa} with the gluelump mass $\Lambda_1=0.87(15)$~GeV from Ref.~\cite{Bali:2003jq}.}
	\label{potentials}
 \end{figure}

For hybrid bound states the binding energies are smaller than $E_B\ll\Lambda_{\rm QCD}$ as they are of the order of small energy fluctuations around the minimum of the hybrid static energies
\begin{align}
V^{(0)}_{\la}(r)&=\lim_{t \rightarrow \infty}\frac{i}{t}\ln \langle{\bf R}, {\bf r};\lambda; t/2 |{\bf R}, {\bf r};\lambda;-t/2\rangle=\Lambda_{B}+V^{(0)}_o+b_{B|\la|} r^2+\cdots=E^{(0)}_{|\la|}(r)\,,\label{Vkappa}
\end{align}
where $E^{(0)}_{0}(r)=E^{(0)}_{\Sigma_u^-}(r)$ and $E^{(0)}_{|\pm 1|}(r)=E^{(0)}_{\Pi_u}(r)$. As for standard quarkonium, we are going to use a fit to the full static energies as the static potential instead of a multipole expanded expression in the middle equality in Eq.~\eqref{Vkappa} in order to increase accuracy. In Fig.~\ref{potentials} we show the lattice data of Ref.~\cite{Juge:2002br} for these static energies and our fitted potentials.

To go beyond the static limit we use that an eigenstate of the full Hamiltonian can be expressed in the basis of eigenstates of the static limit
\begin{align}
|H_n\rangle=\int d^3\bm{r}d^3\bm{R}\sum_{\lambda}\psi^{(n)}_{\lambda}(\bm{R}, \bm{r})|\bm{R}, \bm{r};\lambda\rangle\,.
\end{align}
Using quantum mechanical perturbation theory to incorporate the kinetic operator (see Ref.~\cite{Berwein:2015vca} for a full discussion) one arrives at the coupled Shr\"odinger equations for the hybrid bound states
\begin{align}
\sum_{\la}\left(-\bm{\hat{r}}^{*}_{\lap}\frac{\bnabla^2_r}{m_Q}\bm{\hat{r}}_{\la}+V^{(0)}_{\la}(r)\delta_{\lap\la}\right)\psi^{(n)}_{\la}(\bm{r})={\cal E}_n \psi^{(n)}_{\lap}(\bm{r})\,.\label{scheq}
\end{align}
As in standard quarkonium we use the following short-hand notation for the hybrid quarkonium wave function 
\begin{align}
\psi^{(n)}_{\la}(\bm{r})\equiv \psi_\la^{n j \ell s }(\bm{r})=\psi^\la_{n}(r)\Phi^\la_{^{2s+1}\ell_j}(\theta,\phi)\,,
\end{align}
where we use $(n)$ to denote the set of quantum numbers that define a particular hybrid state. Notice that the hybrid angular wave functions are not an eigenstate of the heavy-quark angular momentum, $\bm{L}^2_{\bar{Q}Q}$, but instead of $\left(\bm{L}_{\bar{Q}Q}+\bm{S}_1\right)^2$, where $\bm{S}_1$ is the spin-$1$ gluelump spin operator~\cite{Berwein:2015vca}, with eigenvalue $\ell(\ell+1)$. To highlight this difference we use a modified spectroscopic notation where $\ell=0,1,2,\dots$ is represented by ${\cal S}$, ${\cal P}$, ${\cal D}$, etc. Moreover the angular wave functions are also eigenstates of $\hat{\bm{r}}\cdot\bm{S}_1$ with eigenvalue $\lambda=-1,0,1$. Due to the projection vectors in each side of the kinetic operator in Eq.~\eqref{scheq} the contributions from the $\Pi_u$ and $\Sigma_u^-$ static potentials are mixed which gives rise to pairs of solutions with the same principal and angular quantum numbers but opposite parity\cite{Berwein:2015vca}. Therefore, the parity should also be specified to single out a specific solution of Eq.~\eqref{scheq}. These two solutions are characterized by different radial wave functions,
\begin{align}                            
\psi^{(n)}_{+}(r)=\left(\begin{array}{c} \psi^{(n)}_0(r) \\ \frac{1}{\sqrt{2}}\psi^{(n)}_+(r) \\ \frac{1}{\sqrt{2}}\psi^{(n)}_+(r) \\ \end{array}\right)\,,\quad
\psi^{(n)}_{-}(r)=\left(\begin{array}{c} 0 \\ \frac{1}{\sqrt{2}}\psi^{(n)}_-(r) \\ -\frac{1}{\sqrt{2}}\psi^{(n)}_-(r) \\ \end{array}\right)\,,\label{rwfs}
\end{align}
with party corresponding to $P=\pm(-1)^{\ell+s}$. The two lowest laying $1^{--}$ hybrid states correspond to $n=1$, $\ell=1$ and $s=0$ and therefore correspond to the first type of solution in Eq.~\eqref{rwfs}.

\subsection{Gluelump overlap with \texorpdfstring{$\bm{B}$}{B}}

The gluelump operator $\bm{G}_{B}$ is a sum of all possible gluonic operators with the same quantum numbers with unknown coefficients
\begin{align}
\bm{G}_{B}=Z^{-1/2}_B g\bm{B}^a+Z^{-1/2}_{D\times E}\left(\bm{D}\times g\bm{E}\right)^a+\cdots\label{glmpexp}
\end{align}
We will follow the hypothesis, proposed in Ref.~\cite{Pineda:2019mhw}, that there is a correlation between the dimensionality of the interpolating operator and the strength of the interpolation, such that higher dimension operators in Eq.~\eqref{glmpexp} are subleading and therefore the series can be truncated at LO. One can estimate the value of $Z_B$ using the normalization of the gluelump operators in Eq.~\eqref{glmpn} to relate it to the value of the gluon condensate
\begin{align}
Z_B=\frac{1}{3}\langle 0|g^2\bm{B}^2|0\rangle=\frac{\pi^2}{3}\langle 0| \frac{\alpha}{\pi}G^a_{\mu\nu}G^{a\mu\nu}|0\rangle=0.251(14)~{\rm GeV}^4\,,\label{zbv}
\end{align} 
with the value taken from Ref.~\cite{Ayala:2020pxq}. We note that the value in Ref.~\cite{Ayala:2020pxq} corresponds to a quenched computation and that the uncertainty in Eq.~\eqref{zbv} does not include an estimation of the uncertainty of this or the truncation of the expansion in Eq.~\eqref{glmpexp}.

\section{LO transitions}\label{s2}

Now we look at the transitions generated by the first operator in Eq.~\eqref{pnrqcd2}. Let ${\cal O}_{\pi}$ denote a generic final light-quark state. The transition amplitude is
\begin{align}
{\cal A}=\langle S_{m}{\cal O}_{\pi}|g{\rm Tr}\left[\S^{\dag}\bm{r}\cdot\bm{E}\,\Oc\right]|H_n\rangle=\frac{1}{3}\sqrt{\frac{T_F}{N_cZ_B}}\langle {\cal O}_{\pi}|g^2\bm{E}\cdot\bm{B}| 0\rangle \int d^3r\sum_{\lambda}\phi^{(m)}(\bm{r})\bm{r}\cdot\bm{\hat{r}}_\lambda\psi^{(n)}_{\lambda}(\bm{r})\,.\label{tt1s1}
\end{align}  
The gluonic operator has quantum numbers $0^{-+}$ and isospin $I=0$, therefore the allowed final light-quark states must match these quantum numbers. Some examples of these states are $\pi^0$, $\eta$, $\eta'$, higher mass $\eta$-like resonances or odd numbers of mesons such as $\pi^0\pi^+\pi^-$ or $\eta\,\pi^+\pi^-$. Selection rules can be derived from the wave functions integral. Since the transition operator is independent of the heavy-quark spin this should be conserved. Furthermore, only the $\lambda=0$ component of the hybrid wave function contributes to the integral because $\bm{r}\cdot\bm{\hat{r}}_\lambda=r\delta_{0\la}$. For $\lambda=0$ the orbital wave function reduces to the usual spherical harmonics and hence $\ell=l$.

If we identify $\Upsilon(10753)$ and $\Upsilon(11020)$ as hybrid quarkonia with $n^1{\cal P}_1$ and $n=1,2$, respectively, then the final quarkonium states must be $h_b(m^1P_1)$. In Table~\ref{lot_aim} we collect the mass difference between these states for the transitions we are going to compute.

\begin{table}[ht!]
\begin{tabular}{|c||c|c|} \hline
 {\rm Mass difference}& $h_b(1^1P_1)$ & $h_b(2^1P_1)$ \\ \hline\hline
$\Upsilon(10753)$ & $854$ & $493$ \\ \hline
$\Upsilon(11020)$ & $1101$ & $740$ \\ \hline
\end{tabular}
\caption{Mass difference for the transitions $H(n^1 {\cal P}_1)\to h_b(m^1P_1)$ in MeV.}
\label{lot_aim}
\end{table}

An explicit computation of the wave function integral can be done using the following expression for the angular wave function:
\begin{align}
\Phi^\la_{^1{\cal P}_1}=\sqrt{\frac{3}{4\pi}}\hat{\bm{r}}^{*}_\la\cdot \hat{e}_{m_j}\frac{\mathbb{1}_2}{\sqrt{2}}\,,\label{awv1p}
\end{align}
which with $\la=0$ also applies to $h_b(m^1P_1)$ states. $\hat{e}_{m_j}$ are the usual polarization vectors with $m_j$ the eigenvalue of $J_3$. The matrix elements for the production of a single $\pi^0,\,\eta$ or $\eta'$ can be obtained from the $U(1)_A$ anomaly using a mixing scheme~\cite{Bramon:1997va,Feldmann:1998vh,Kroll:2005sd,Escribano:2020jdy}. We will use the mixing scheme of Ref.~\cite{Kroll:2005sd} which we briefly summarize in Appendix~\ref{Ap:krollm}. Further kinematically allowed final states do exist, but are left for a future work. We arrive at the following transition amplitude:
\begin{align}
{\cal A}_{\left(n^1{\cal P}_1\right)\to \left(m^{1}P_1\right) P}=\delta_{m_jm_{j'}}\frac{4\pi^2}{3}\sqrt{\frac{T_F}{N_c Z_B}}\omega_P \langle m|r|n,\,0\rangle\quad P=\pi^0,\,\eta,\,\eta'\,,\label{smfe:e1}
\end{align}
with $\omega_P$ given in Eqs.~\eqref{ompi0}-\eqref{ometap} and we have used the following short-hand notation for the remaining integration of the radial wave functions:
\begin{align}
\langle m|r^p|n,\lambda\rangle=\int_0^\infty dr\,r^{2+p}\phi_{m}(r)\psi_n^{\lambda}(r)\,,
\end{align}
which we evaluate numerically.

The corresponding width is obtained using Eq.~\eqref{smfe:e1} into Eq.~\eqref{dw1p} and averaging over initial polarizations and summing over the final ones. For the transitions in table~\ref{lot_aim} with $\pi^0,\,\eta$ or $\eta'$ in the final state we obtain the following results:
\begin{align}
&\Gamma_{\Upsilon(10753)\to h_b(1P)\pi^0}=2.57(\pm 1.03)_{\rm m.e.}(\pm 0.14)_{Z_B}(\pm 0.16)_{\omega_{\pi^0}}~{\rm keV}\,,\label{s2:t1}\\
&\Gamma_{\Upsilon(10753)\to h_b(1P)\eta}=2.29(\pm 0.92)_{\rm m.e.}(\pm 0.13)_{Z_B}(\pm 0.08)_{\omega_{\eta}}~{\rm MeV}\,,\label{s2:t2}\\
&\Gamma_{\Upsilon(10753)\to h_b(2P)\pi^0}=0.168(\pm 0.067)_{\rm m.e.}(\pm 0.009)_{Z_B}(\pm 0.010)_{\omega_{\pi^0}}~{\rm keV}\,,\label{s2:t3}\\
&\Gamma_{\Upsilon(11020)\to h_b(1P)\pi^0}=2.04(\pm 0.82)_{\rm m.e.}(\pm 0.11)_{Z_B}(\pm 0.13)_{\omega_{\pi^0}}~{\rm keV}\,,\label{s2:t4}\\
&\Gamma_{\Upsilon(11020)\to h_b(1P)\eta}=2.04(\pm 0.81)_{\rm m.e.}(\pm 0.11)_{Z_B}(\pm 0.07)_{\omega_{\eta}}~{\rm MeV}\,,\label{s2:t5}\\
&\Gamma_{\Upsilon(11020)\to h_b(1P)\eta'}=9.23(\pm 3.69)_{\rm m.e.}(\pm 0.51)_{Z_B}(\pm 0.39)_{\omega_{\eta'}}~{\rm MeV}\,,\label{s2:t6}\\
&\Gamma_{\Upsilon(11020)\to h_b(2P)\pi^0}=0.104(\pm 0.042)_{\rm m.e.}(\pm 0.006)_{Z_B}(\pm 0.006)_{\omega_{\pi^0}}~{\rm keV}\,,\label{s2:t7}\\
&\Gamma_{\Upsilon(11020)\to h_b(2P)\eta}=81.8(\pm 32.7)_{\rm m.e.}(\pm 4.6)_{Z_B}(\pm 2.7)_{\omega_{\eta}}~{\rm keV}\,.\label{s2:t8}
\end{align}

We estimate the uncertainty from using the multipole expansion (${\rm m.e.}$) as corrections of ${\cal O}\left(\Lambda^2_{\rm QCD}r^2\right)$. In the multipole expansion, the heavy-quark distance scales as $1/r \sim m_Q v\gg \Lambda_{\rm QCD}$, with $v$ the heavy-quark pair relative velocity. On the other, hand the adiabatic expansion between heavy and light degrees of freedom requires $\Lambda_{\rm QCD}\gg E_b\sim m_Q v^2$, with $E_b$ the binding energy of the heavy quarks~\cite{Brambilla:2017uyf}. A scaling of $\Lambda_{\rm QCD}$ consistent with these two constraints is $\Lambda_{\rm QCD}\sim m_Q v^{3/2}$. Thus ${\cal O}\left(\Lambda^2_{\rm QCD}r^2\right)\sim v\sim 0.4$ for the states we consider. The uncertainties labeled as $Z_B$ and $\omega_P$, $P=\pi^0,\,\eta,\,\eta'$ are just the standard propagation of the uncertainty of these quantities in Eq.~\eqref{zbv} and Eqs.~\eqref{ompi0num}-\eqref{ometapnum}.  

\section{NLO transitions}\label{s3}

Now, we study the transitions generated by the $1/m_Q$ order operator of the Lagrangian in Eq.~\eqref{pnrqcd2}. The transition amplitude is as follows:
\begin{align}
{\cal A}=\langle S_{m}{\cal O}_{\pi\pi}|\frac{gc_F}{m_Q}{\rm Tr}\left[\S^{\dag}(\bm{S}_1-\bm{S}_2)\cdot\bm{B}\,\Oc\right]|H_n\rangle=\frac{gc_F}{3m_Q}\sqrt{\frac{T_F}{N_cZ_B}}\langle {\cal O}_{\pi\pi}|\bm{B}^2| 0\rangle \int d^3r\sum_{\lambda}\phi^{(m)}(\bm{r})(\bm{S}_1-\bm{S}_2)\cdot\bm{\hat{r}}_\lambda\psi^{(n)}_{\lambda}(\bm{r})\,,\label{tt2s1}
\end{align}  
where ${\cal O}_{\pi\pi}$ denotes generically the final light-quark meson state. Since $\bm{B}^2$ has quantum numbers $0^{++}$ and $I=0$, these must be the quantum numbers of the light-quark final state. Such states are, for instance $\pi^+\pi^-$, $K^+K^-$, pairs of $\pi^0$ or $\eta$ as well as $f_0$ resonances up to the invariant mass allowed by the specific initial and final heavy-quark states. The heavy-quark spin structure of the operator requires the transitions to be between singlet and triplet states. Therefore, if the initial state is a spin singlet hybrid, as per our assignations of the $\Upsilon(10753)$ and $\Upsilon(11020)$, the final states must be spin triplet quarkonium states. Furthermore the total $J^{PC}$ must be conserved, therefore the final quarkonium states can only be $\Upsilon(m^3S_1)$ or $\Upsilon(m^3D_1)$. We will only consider the first case since $D$-wave bottomonium states have not yet been observed experimentally. In Table~\ref{nlot_aim} we collect the mass differences between the initial and final heavy-quark states for the  transitions $H(n^1 {\cal P}_1)\to \Upsilon(m^3S_1)$ that we will consider.

\begin{table}[ht!]
\begin{tabular}{|c||c|c|c|} \hline
{\rm Mass difference} & $\Upsilon(1^3S_1)$ & $\Upsilon(2^3S_1)$ & $\Upsilon(3^3S_1)$ \\ \hline\hline
$\Upsilon(10753)$ & $1293$ & $730$ & $398$ \\ \hline
$\Upsilon(11020)$ & $1540$ & $977$ & $645$ \\ \hline
\end{tabular}
\caption{Mass difference for the transitions $H(n^1 {\cal P}_1)\to \Upsilon(m^3S_1)$ in MeV.}
\label{nlot_aim}
\end{table}

The integral of the angular wave function in Eq.~\eqref{pnrqcd2} can be carried out using the wave function in Eq.~\eqref{awv1p} for the $^1 {\cal P}_1$ hybrid state and
\begin{align}
\Phi^0_{^3S_1}=\frac{1}{\sqrt{4\pi}}\frac{\mathbb{1}_2}{\sqrt{2}}\,,
\end{align}
for the $^3S_1$ standard quarkonium. 

The transition widths for the amplitude in Eq.~\eqref{tt2s1} with two pions or kaons in the final state reads as
\begin{align}
{\cal A}_{\left(n^1{\cal P}_1\right)\to \left(m^{3}S_1\right) P^+P^-}=\delta_{m_jm_{j'}}\frac{8\pi^2c_F}{3\beta_0m_Q}\sqrt{\frac{T_F}{3N_c Z_B}}\left(\sum_\la\langle m|1|n,\la\rangle\right)F_P(s,\cos\theta)\quad P=\pi,\,K\,,\label{smfe:e2}
\end{align}
with the form factor $F_P(s,\cos\theta)$ that encodes the production of two pions or kaons by the gluonic operator $\bm{B}^2$ defined in Eq.~\eqref{Ap:om:e11}. The definitions of $s$ and $\theta$ can be found in Appendix~\ref{Ap:mvw}. In Appendix~\ref{Ap:om} we give a dispersive representation of $F_P(s,\cos\theta)$. Our approach consists in a coupled Muskhelishvili-Omn\`es equations as in Refs.~\cite{Donoghue:1990xh,Moussallam:1999aq,Celis:2013xja,Chen:2016mjn}. However, unlike those references our matrix element contains not only an $S$-wave piece but also $D$-wave one. We have extended the coupled Muskhelishvili-Omn\`es approach to the $D$-wave final state interactions for the first time. We use the parametrizations of the $\pi\pi\to \pi\pi$ and $\pi\pi\to K\bar{K}$ partial waves from Refs.~\cite{GarciaMartin:2011cn,Pelaez:2018qny}, which to our knowledge are the most accurate currently available. For the numerical solution of the coupled Muskhelishvili-Omn\`es equations we use the techniques of Refs.~\cite{Moussallam:1999aq,Descotes-Genon:2000pfd}.

\begin{figure}[ht!]
\begin{tabular}{ccc}
\includegraphics[width=.49\textwidth]{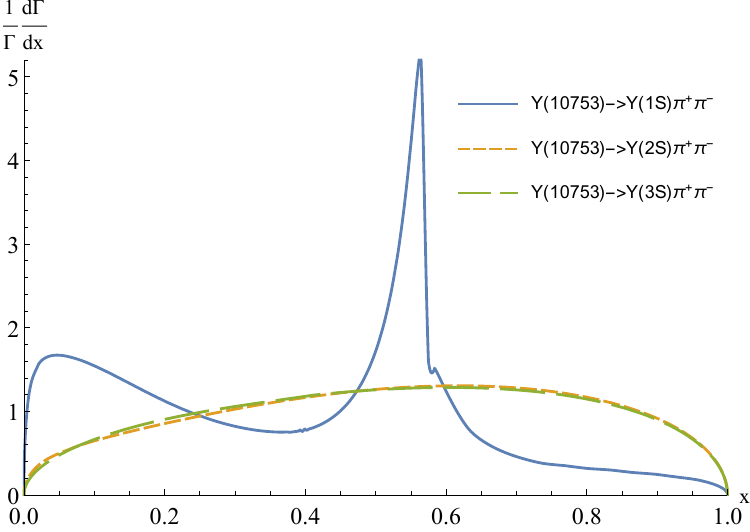} & \includegraphics[width=.49\textwidth]{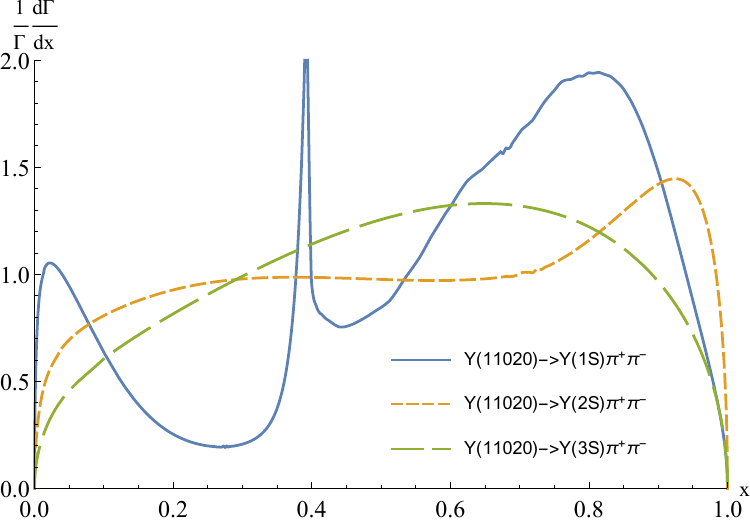}   \\
\end{tabular}
\caption{Normalized differential width for the transitions $H(n^1{\cal P}_1)\to \Upsilon(m^3S_1)\pi^+\pi^-$. The variable $x$ is defined as $x=(s-4m^2_{\pi})/(m_{H(n{\cal P})}-m_{\Upsilon(mS)}-4m^2_{\pi})$.}
	\label{lshpp}
\end{figure}

\begin{figure}[ht!]
\centerline{\includegraphics[width=.49\textwidth]{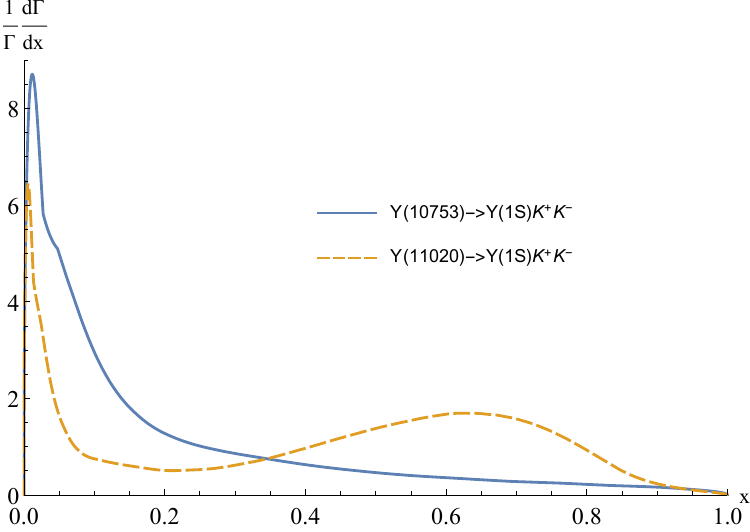}}
\caption{Normalized differential width for the transitions $H(n^1{\cal P}_1)\to \Upsilon(1^3S_1)K^+K^-$. The variable $x$ is defined as $x=(s-4m^2_{K})/(m_{H(n{\cal P})}-m_{\Upsilon(1S)}-4m^2_{K})$.}
	\label{lshkk}
\end{figure}

The transition differential widths can be computed from the amplitude in Eq.~\eqref{smfe:e2} by decomposing it into partial waves and using Eq.~\eqref{Ap:mvw:e4}. In Figs.~\ref{lshpp} and \ref{lshkk} we plot the normalized differential decay widths for the transitions with $\pi^+\pi^-$ and $K^+K^-$ in the final states, respectively. The normalized differential decay widths are independent of the heavy-quark matrix elements, therefore the line shapes are a result of the pion and kaon rescattering as well as the phase space dependence. The prominent features of the $\Upsilon(10753)\to \Upsilon(1S)\pi^+\pi^-$, $\Upsilon(11020)\to \Upsilon(1S)\pi^+\pi^-$, $\Upsilon(10753)\to \Upsilon(1S)K^+K^-$ and $\Upsilon(11020)\to \Upsilon(1S)K^+K^-$ line shapes makes these good observables to study experimentally.


Integrating the differential transition width over the kinematically allowed range of $s$ we obtain the following transition widths:
\begin{align}
\Gamma_{\Upsilon(10753)\to \Upsilon(1S)\pi^+\pi^-}&=43.4(\pm 17.3)_{\rm m.e.}(\pm 2.4)_{Z_B}(\pm 8.6)_{\alpha_s}(^{+0.5}_{-0.0})_{\kappa}~{\rm keV}\,,\label{s3:e1}\\
\Gamma_{\Upsilon(10753)\to \Upsilon(2S)\pi^+\pi^-}&=2.75(\pm 1.10)_{\rm m.e.}(\pm 0.15)_{Z_B}(\pm 0.55)_{\alpha_s}(^{+0.13}_{-0.12})_{\kappa}~{\rm keV}\,,\label{s3:e2}\\
\Gamma_{\Upsilon(10753)\to \Upsilon(3S)\pi^+\pi^-}&=0.98(\pm 0.39)_{\rm m.e.}(\pm 0.05)_{Z_B}(\pm 0.19)_{\alpha_s}(\pm 0.03)_{\kappa}~{\rm eV}\,,\label{s3:e3}\\
\Gamma_{\Upsilon(11020)\to \Upsilon(1S)\pi^+\pi^-}&=99.1(\pm 39.6)_{\rm m.e.}(\pm 5.5)_{Z_B}(\pm 19.7)_{\alpha_s}(^{+26.3}_{-21.8})_{\kappa}~{\rm keV}\,,\label{s3:e4}\\
\Gamma_{\Upsilon(11020)\to \Upsilon(2S)\pi^+\pi^-}&=3.96(\pm 1.58)_{\rm m.e.}(\pm 0.22)_{Z_B}(\pm 0.70)_{\alpha_s}(^{-0.16}_{+0.17})_{\kappa}~{\rm keV}\,,\label{s3:e5}\\
\Gamma_{\Upsilon(11020)\to \Upsilon(3S)\pi^+\pi^-}&=1.33(\pm 0.53)_{\rm m.e.}(\pm 0.07)_{Z_B}(\pm 0.27)_{\alpha_s}(\pm 0.02)_{\kappa}~{\rm keV}\,,\label{s3:e6}\\
\Gamma_{\Upsilon(10753)\to \Upsilon(1S)K^+K^-}&=3.98(\pm 1.59)_{\rm m.e.}(\pm 0.22)_{Z_B}(\pm 0.79)_{\alpha_s}(^{-0.50}_{+0.67})_{\kappa}~{\rm keV}\,,\label{s3:e7}\\
\Gamma_{\Upsilon(11020)\to \Upsilon(1S)K^+K^-}&=5.93(\pm 2.37)_{\rm m.e.}(\pm 0.33)_{Z_B}(\pm 1.18)_{\alpha_s}(^{+1.75}_{-1.18})_{\kappa}~{\rm keV}\,.\label{s3:e8}
\end{align}
We have used the renormalization group improved expression of $c_F(1~{\rm GeV})=0.879$ up to next-to-leading logarithmic order, with the values $\alpha_s(1~{\rm GeV})$ = 0.4798 and $\alpha_s(m_b)=0.214820$ and $m_b=4.885$~GeV~\cite{Peset:2018ria}. The values of $\alpha_s$ were computed using the \texttt{RunDec} Mathematica package~\cite{Chetyrkin:2000yt}. 

The uncertainties are labeled by their source of origin. The subscript ${\rm m.e.}$ denotes the uncertainty stemming from the use of the multipole expansion. As in the previous section these are estimated as corrections of ${\cal O}\left(\Lambda^2_{\rm QCD}r^2\right)\sim v\sim 0.4$. The uncertainty labeled as $Z_B$ is just the standard propagation of the uncertainty in Eq.~\eqref{zbv}. The uncertainty of the dispersive parametrization of the form factors is dominated by the uncertainties in the chiral representation in Eq.~\eqref{Ap:om:e9} to which it is matched to in order to determine the subtraction polynomials. There are two of these sources of uncertainty: the first one from neglecting the anomalous dimension $\gamma_i$ and truncating $\beta(\alpha_s)/\alpha_s$ at LO in Eq.~\eqref{Ap:om:e13} which we label with the subscript $\alpha_s$ and are of order $\alpha_s(m_c)/(4\pi)$~\cite{Chivukula:1989ds}. The second one is associated to the value of the parameter $\kappa$ in Eq.~\eqref{Ap:om:e12} which affects the form factors asymmetrically. Other sources of uncertainty for the form factors, such as the parametrization of $\pi\pi\to \pi\pi$ and $\pi\pi\to K\bar{K}$ phase shifts are negligible in front of the other sources.

Experimental values for some of the transition widths in Eqs.~\eqref{s3:e1}-\eqref{s3:e8} are available. In Ref.~\cite{Abdesselam:2019gth} the widths for the transitions $\Upsilon(11020)\to \Upsilon(nS)\pi^+\pi^-$, $n=1,2,3$ are given normalized to $\Gamma_{e^+e^-}/\Gamma_{\rm total}$. The latter can be obtained from the PDG average of Refs.~\cite{Besson:1984bd,Lovelock:1985nb}. The values, summing the uncertainties quadratically, are as follows:
\begin{align}
\Gamma^{\rm exp}_{\Upsilon(11020)\to \Upsilon(1S)\pi^+\pi^-}&=85^{+33}_{-36}~{\rm keV}\,,\label{s3:e9}\\
\Gamma^{\rm exp}_{\Upsilon(11020)\to \Upsilon(2S)\pi^+\pi^-}&=120^{+105}_{-107}~{\rm keV}\,,\label{s3:e10}\\
\Gamma^{\rm exp}_{\Upsilon(11020)\to \Upsilon(3S)\pi^+\pi^-}&=61^{+37}_{-38}~{\rm keV}\,.\label{s3:e11}
\end{align}
It is remarkable the agreement between our theoretical value for the width of $\Upsilon(11020)\to \Upsilon(1S)\pi^+\pi^-$ in Eq.~\eqref{s3:e4} with the experimental value in Eq.~\eqref{s3:e9}. Nevertheless, one should be cautious considering the significant uncertainties on both theoretical and experimental values. On the other hand, the values for the widths of the $\Upsilon(11020)\to \Upsilon(nS)\pi^+\pi^-$, $n=2,3$ transitions are not compatible with our theoretical values in Eqs.~\eqref{s3:e5} and \eqref{s3:e6}. However, it should be noted that the values in Eqs.~\eqref{s3:e9}-\eqref{s3:e11} correspond to the range of solutions from different fits and not $1\sigma$ intervals. Therefore, it is still possible that future experimental studies produce a closer result to our predictions.

\section{Ratios}\label{s4}

Our results for the $\Upsilon(10753)$ and $\Upsilon(11020)$ transition widths into standard quarkonium have overall large uncertainties as discussed in Secs.~\ref{s2} and \ref{s3}. We can obtain more precise predictions if we consider various ratios of transition widths where some uncertainties cancel out. 

Let us consider the following ratios of the LO transitions with the same initial and final heavy-quark states
\begin{align}
&\frac{\Gamma_{\Upsilon(10753)\to h_b(1P)\eta}}{\Gamma_{\Upsilon(10753)\to h_b(1P)\pi^0}}=891(\pm 64)_{\omega}\,,\label{s4a:e1}\\
&\frac{\Gamma_{\Upsilon(11020)\to h_b(1P)\eta}}{\Gamma_{\Upsilon(11020)\to h_b(1P)\pi^0}}=1001(\pm 72)_{\omega} \,,\\
&\frac{\Gamma_{\Upsilon(11020)\to h_b(1P)\eta'}}{\Gamma_{\Upsilon(11020)\to h_b(1P)\eta}}=4.52(\pm 0.25)_{\omega}\,,\\
&\frac{\Gamma_{\Upsilon(11020)\to h_b(2P)\eta}}{\Gamma_{\Upsilon(11020)\to h_b(2P)\pi^0}}=786(\pm 52)_{\omega}\,.\label{s4a:e2}
\end{align}
These depend only on phase space factors and the values of the gluonic matrix elements $\omega_P$. Therefore, these ratios are free of the uncertainties associated to the multipole expansion or the value of $Z_B$. On the other hand the value of these ratios is mainly a test of the $\pi^0-\eta-\eta'$ mixing scheme of Ref.~\cite{Kroll:2005sd}. Nevertheless, these ratios rely on the factorization of the gluonic matrix elements from the heavy-quark physics, which in our approach is the result of the multipole expansion. Hence one can test this factorization from experimental values of the ratios in Eqs.~\eqref{s4a:e1}-\eqref{s4a:e2}. One can test more directly the validity of the multipole expansion for $\Upsilon(10753)$ and $\Upsilon(11020)$ by considering the ratios with the same final state light-quark meson but different final standard quarkonium,
\begin{align}
&\frac{\Gamma_{\Upsilon(10753)\to h_b(1P)\pi^0}}{\Gamma_{\Upsilon(10753)\to h_b(2P)\pi^0}}=15.2(\pm 8.1)_{\rm m.e.}\,,\\
&\frac{\Gamma_{\Upsilon(11020)\to h_b(1P)\pi^0}}{\Gamma_{\Upsilon(11020)\to h_b(2P)\pi^0}}=19.6(\pm 11.1)_{\rm m.e.}\,,\\
&\frac{\Gamma_{\Upsilon(11020)\to h_b(1P)\eta}}{\Gamma_{\Upsilon(11020)\to h_b(2P)\eta}}=24.9(\pm 14.1)_{\rm m.e.}\,,
\end{align}
since in this case the factors $\omega_P$ and $Z_B$ cancel out.

Next we consider ratios of the NLO transitions in Eqs.~\eqref{s3:e1}-\eqref{s3:e8}. Any ratio of these transition widths is independent of the value of $Z_B$. Furthermore, the uncertainty due to the truncation of the $\beta$ function in Eq.~\eqref{Ap:om:e13} also cancels out and the only remaining uncertainty of order $\alpha_s(m_c)/(4\pi)$ comes from neglecting the anomalous dimension $\gamma_i$. As we have seen for the LO transitions, the ratios of transitions with the same initial and final heavy-quark states are independent of the heavy-quark matrix elements and therefore the uncertainties related to the multipole expansion are not present. We obtain:
\begin{align}
&\frac{\Gamma_{\Upsilon(10753)\to \Upsilon (1S)\pi^+\pi^-}}{\Gamma_{\Upsilon(10753)\to \Upsilon (1S)K^+K^-}}=10.9(\pm 2.2)_{\alpha_s}(^{+1.7}_{-1.5})_{\kappa}\,,\label{s4a:e3}\\
&\frac{\Gamma_{\Upsilon(11020)\to \Upsilon (1S)\pi^+\pi^-}}{\Gamma_{\Upsilon(11020)\to \Upsilon (1S)K^+K^-}}=16.7(\pm 3.3)_{\alpha_s}(^{+4.4}_{-3.7})_{\kappa}\,.\label{s4a:e4}
\end{align}
The values of the ratios in Eqs.~\eqref{s4a:e3} and \eqref{s4a:e4} depend mainly on the dispersive representation of the form factors and therefore can be considered a test of it. However, this is dependent on the factorization of the gluonic matrix elements in an analogous manner to our discussion for the ratios of the LO transitions in Eqs.~\eqref{s4a:e1}-\eqref{s4a:e2}. One can consider $26$ additional ratios among the widths in Eqs.~\eqref{s3:e1}-\eqref{s3:e8} with similar cancellation of uncertainties. We are going to focus on the ones with the same initial state and different final one, since those are most interesting in order to learn about the nature of $\Upsilon(10753)$ and $\Upsilon(11020)$. These ratios are as follows
\begin{align}
&\frac{\Gamma_{\Upsilon(11020)\to \Upsilon (1S)\pi^+\pi^-}}{\Gamma_{\Upsilon(10753)\to \Upsilon (1S)\pi^+\pi^-}}=2.28(\pm 0.91)_{\rm m.e.}(\pm 0.46)_{\alpha_s}(^{+0.57}_{-0.51})_{\kappa}\,,\label{s4a:e5}\\
&\frac{\Gamma_{\Upsilon(11020)\to \Upsilon (2S)\pi^+\pi^-}}{\Gamma_{\Upsilon(10753)\to \Upsilon (2S)\pi^+\pi^-}}=1.44(\pm 0.57)_{\rm m.e.}(\pm 0.29)_{\alpha_s}(^{-0.06}_{+0.07})_{\kappa}\,,\label{s4a:e6}\\
&\frac{\Gamma_{\Upsilon(11020)\to \Upsilon (3S)\pi^+\pi^-}}{\Gamma_{\Upsilon(10753)\to \Upsilon (3S)\pi^+\pi^-}}=1.36(\pm 0.54)_{\rm m.e.}(\pm 0.27)_{\alpha_s}(\mp 0.02)_{\kappa}\times 10^{3}\,,\label{s4a:e7}\\
&\frac{\Gamma_{\Upsilon(11020)\to \Upsilon (1S)K^+K^-}}{\Gamma_{\Upsilon(10753)\to \Upsilon (1S)K^+K^-}}=1.49(\pm 0.59)_{\rm m.e.}(\pm 0.30)_{\alpha_s}(^{+0.72}_{-0.47})_{\kappa}\,.\label{s4a:e8}
\end{align}
The uncertainty related to the value of $\kappa$ is reduced in the ratios in Eqs.~\eqref{s4a:e6} and \eqref{s4a:e7} but enhanced in the ratios in Eqs.~\eqref{s4a:e4} and \eqref{s4a:e8}. This behavior can be traced to the similarity and difference, respectively, of the normalized spectra in Figs.~\ref{lshpp} and \ref{lshkk}.

Finally, one could consider the ratios of LO and NLO transition widths. These are independent of the value $Z_B$, however one needs to add the uncertainties of the two gluonic operator matrix elements and the multipole expansion uncertainties of the heavy-quark matrix elements, which together make the relative uncertainty of these ratios larger than the ones of the transition widths themselves.

\section{Comparison with semi-inclusive transition widths}\label{s:siw}

\begin{figure}[ht!]
   \centerline{\includegraphics[width=.3\textwidth]{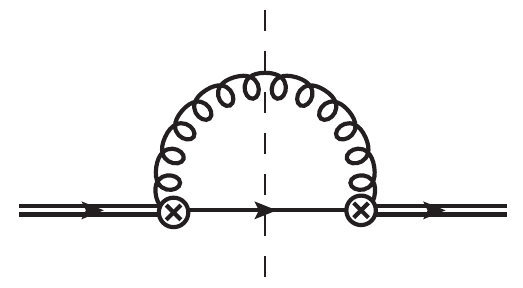}}
	\caption{The single and double lines represent quarkonia in singlet and octet states respectively. The curly line stands for a gluon. The imaginary part of this self-energy diagram, which can be obtained by cutting the diagram by the dashed line, produces the semi-inclusive width associated to the transition from $H_n$ to $S_m$ states with any other light-quark hadrons in the final state. Note that the spectator gluons forming the $H_n$ state are not displayed.}
	\label{tptf}
 \end{figure}

When the energy gap between a hybrid and a standard quarkonium state is large, the gluon emitted by the heavy quarks in the transition from an octet to a singlet state can be considered perturbative and semi-inclusive decay widths can be computed~\cite{Oncala:2017hop}. These semi-inclusive decay widths correspond to the expected value of the hybrid states of the imaginary part of the diagram in Fig.~\ref{tptf}. The vertices in the diagram can be either of the operators in the Lagrangian in Eq.~\eqref{pnrqcd2}. This computation was carried out in Ref.~\cite{Oncala:2017hop} for the LO term of Eq.~\eqref{pnrqcd2}, here we reproduce it and extend it to the NLO operator. We are going to consider only transitions for which $\alpha_s(E_n-E_m)\leq0.5$ with $E_n-E_m$ the energy difference between the hybrid and standard quarkonium. We will compare the semi-inclusive transition width values to our results in Secs.~\ref{s2} and \ref{s3}.

For the LO operator the semi-inclusive width is
\begin{align}
\Gamma^{\rm LO}_{H_m\to S_n}=\frac{4\alpha_s}{3}\frac{T_F}{N_c} (E_n-E_m)^3\langle\psi_m|(\hat{\bm{r}}_\la^*)^k\bm{r}^i|\phi_n\rangle\langle \phi_n|\bm{r}^j\hat{\bm{r}}_\la^k|\psi_m\rangle\,.\label{siw2}
\end{align}
The expected value for transitions from $n^1{\cal P}_1$ to $m^1P_1$
\begin{align}
&\langle n^1{\cal P}_1|(\hat{\bm{r}}_\la^*)^k\bm{r}^i|m^1P_1\rangle\langle m^1P_1|\bm{r}^j\hat{\bm{r}}_\la^k|n^1{\cal P}_1\rangle=\frac{1}{5}\left(3\langle m|r|n,\,0 \rangle^2+2\sqrt{2}\langle m|r|n,\,0 \rangle\langle m|r|n,+1 \rangle+4\langle m|r|n,+1 \rangle^2\right)\,.\label{siw3}
\end{align}
Using Eq.~\eqref{siw3} in Eq.~\eqref{siw2} and taking $\alpha_s$ at the scale $E_n-E_m$ using the \texttt{RunDec} Mathematica package~\cite{Chetyrkin:2000yt} we compute the values for the semi-inclusive widths corresponding to the transitions in Table~\ref{lot_aim}. Only one transition has a large enough energy gap
\begin{align}
&\Gamma^{\rm LO}_{\Upsilon(11020)\to h_b(1P)}=20(\pm 9)_{\alpha_s}~{\rm MeV}\,.\label{s:siw:e1}
\end{align}
The sum of Eqs.~\eqref{s2:t3}-\eqref{s2:t5}, with the uncertainties added in quadrature, is $\Gamma=11\pm 4$~MeV. This value is compatible with the result in Eq.~\eqref{s:siw:e1}, albeit with smaller central value, which might indicate that other final states, such as $\pi^0\pi^+\pi^-$ and $\eta\pi^+\pi^-$, can have transition widths of similar size to the ones we computed.

The semi-inclusive decay width mediated by the NLO operator in the Lagrangian in Eq.~\eqref{pnrqcd2} is
\begin{align}
\Gamma^{\rm NLO}_{H_m \to S_n}=\frac{4c^2_F\alpha_s}{3m^2_Q}\frac{T_F}{N_c}\left(E_n-E_m\right)^3\langle\psi_m|(\hat{\bm{r}}_\la^*)^k\left(\bm{S}_1-\bm{S}_2\right)^i|\phi_n\rangle\langle \phi_n|\left(\bm{S}_1-\bm{S}_2\right)^i\hat{\bm{r}}_\la^k|\psi_m\rangle\,.
\end{align}
The matrix element for transitions from $n^1{\cal P}_1$ to $m^3S_1$ reads
\begin{align}
\langle n^1{\cal P}_1|(\hat{\bm{r}}_\la^*)^k\left(\bm{S}_1-\bm{S}_2\right)^i|m^3S_1\rangle\langle m^3S_1|\left(\bm{S}_1-\bm{S}_2\right)^i\hat{\bm{r}}_\la^k|n^1{\cal P}_1\rangle=\left(\sum_{\la}\langle m|1|n,\,\lambda\rangle\right)^2\,.
\end{align}
In this case three transitions from Table~\ref{nlot_aim} have large enough energy gaps
\begin{align}
&\Gamma^{\rm NLO}_{\Upsilon(10753)\to \Upsilon (1S)}=9.7(\pm 3.8)_{\alpha_s}~{\rm MeV}\,,\label{s:siw:e2}\\
&\Gamma^{\rm NLO}_{\Upsilon(11020)\to \Upsilon (1S)}=7.3(\pm 2.5)_{\alpha_s}~{\rm MeV}\,,\label{s:siw:e3}\\
&\Gamma^{\rm NLO}_{\Upsilon(11020)\to \Upsilon (2S)}=1.1(\pm 0.5)_{\alpha_s}~{\rm MeV}\,.\label{s:siw:e4}
\end{align}
All of these three widths are much larger than the sum of the channels that we have computed for these transitions; the first two, Eqs.~\eqref{s:siw:e2} and \eqref{s:siw:e3}, by $2$ orders of magnitude and the last one, Eq.~\eqref{s:siw:e4}, by $3$ orders of magnitude. Therefore, for the  transitions in Eqs.~\eqref{s:siw:e2}-\eqref{s:siw:e4} we expect large contributions from light-quark final states different from the ones considered here, such as $4\pi$, $\eta\eta$ and $f_0$ resonances.

Finally it is interesting to notice that the sum of semi-inclusive widths for $\Gamma^{\rm LO+NLO}_{\Upsilon(11020)}=28.4\pm 9.4$~MeV is compatible with the experimental value of the total width $\Gamma^{\rm exp}_{\Upsilon(11020)}=24^{+8}_{-6}$~MeV. This is a strong indication that the transitions to $\Upsilon (1S)$, $\Upsilon (2S)$ and $ h_b(1P)$ are the main decay channels for $\Upsilon(11020)$.

\section{Conclusions}\label{s5}

We have computed the transition widths of $\Upsilon(10753)$ and $\Upsilon(11020)$ into standard quarkonium states and light-quark mesons using nonrelativistic EFT. We have worked under the assumption that these two states are the first two lowest laying $1^{--}$ hybrid bottomonium states. The hybrid quarkonium states are heavy-quark-antiquark bound states around the minima of the static energies computed in the quenched approximation in lattice QCD~\cite{Juge:2002br,Capitani:2018rox}. Although the spectrum of quark-antiquark static energies is not known with dynamical light quarks, the results from Ref.~\cite{Bali:2000vr} for the two lowest static energies show negligible difference to the quenched approximation ones. Therefore, it is plausible that the hybrid bottomonium states used in our approach are a good approximation of a more general isospin $I=0$ exotic quarkonium state mixing a nontrivial gluonic component with a light-quark-antiquark pair component.

Hybrid quarkonium states can be described in an EFT setting that incorporates the heavy-quark mass expansion and an adiabatic expansion between the heavy quark and light degrees of freedom~\cite{Berwein:2015vca,Oncala:2017hop}. Since the EFT coincides with the Born-Oppenheimer approximation at LO it is sometimes referred to as Born-Oppenheimer EFT~\cite{Brambilla:2017uyf}. In this EFT framework the two lowest laying $1^{--}$ hybrid states correspond to the ground state and the first radial excitation of the coupled $\Sigma_u^--\Pi_u$ static energies, with $\ell=1$, negative parity and singlet heavy-quark spin~\cite{Berwein:2015vca}. Nevertheless, in this paper we do not formally perform the adiabatic expansion at the Lagrangian level since we are interested in the transitions to standard quarkonium. The standard quarkonium states are the bound states over the ground state static energy $\Sigma_g^+$. We have used the lattice data for the static energies from Ref.~\cite{Juge:2002br}. 

To study the transitions we work in weakly coupled pNRQCD~\cite{Pineda:1997bj,Brambilla:1999xf} an EFT incorporating the heavy-quark mass and multipole expansions. Since it is doubtful that the multipole expansion can be employed in the hybrid charmonium sector, we have restricted ourselves to the bottomonium one. In the multipole expansion, the transition amplitudes factorize into a the heavy-quark matrix element and a gluonic matrix element that creates the final light-quark states. We have studied the transitions generated by the singlet-octet field couplings at NLO in the multipole or heavy-quark mass expansions that can be found in the Lagrangian in Eq.~\eqref{pnrqcd2}. We have found that the LO transition operator generates transitions from $\Upsilon(10753)$ and $\Upsilon(11020)$ to $h_b(mP)$  quarkonium with emission of light-quark mesons in a $0^{-+}$ state. The NLO transition operator, suppressed by the heavy-quark mass, generates transitions from $\Upsilon(10753)$ and $\Upsilon(11020)$ to $\Upsilon(mS)$ with $0^{++}$ light-quark meson states.

In the case of the LO transitions we have computed the widths for transitions with $\pi^0$, $\eta$, $\eta'$ in the final state. The gluonic production matrix elements are obtained employing the $U(1)_A$ anomaly and the mixing scheme from Ref.~\cite{Kroll:2005sd}. The values of the transition widths can be found in Eqs.~\eqref{s2:t1}-\eqref{s2:t8}. Our estimate for the uncertainties of these widths are large and dominated by the multipole expansion corrections. The gluonic matrix elements have small uncertainties except for $\pi^0$ production due to this matrix element being proportional to the difference of the $u$ and $d$ quark masses.

For the NLO transitions we consider the light-quark final states $\pi^+\pi^-$ and $K^+K^-$. The corresponding production matrix elements are obtained through a dispersive representation similar to the one in Refs.~\cite{Donoghue:1990xh,Moussallam:1999aq,Celis:2013xja}. This consists of two coupled Muskhelishvili-Omn\`es integral equations for the $\pi^+\pi^-$ and $K^+K^-$ channels. The $T$-matrix inputs are taken from Refs.~\cite{GarciaMartin:2011cn,Pelaez:2018qny}, which as a whole provide accurate results up to $\sqrt{s}=1.42$~GeV, with $s$ the squared sum of the momenta of $\pi^+\pi^-$ or $K^+K^-$. The numerical solution of the integral equations is obtained using the techniques of Refs.~\cite{Moussallam:1999aq,Descotes-Genon:2000pfd}. Since the gluonic operator contains both an $S$- and $D$-wave pieces we have solved the coupled Muskhelishvili-Omn\`es equations for both waves. For the $S$- wave case we reproduce the results in the literature. The results for the $D$ wave are presented here for the first time. Our results are plotted in Figs.~\ref{omatrixs} and \ref{omatrixd}. The subtraction polynomials are obtained by matching to a chiral representation with the low-energy constants partially determined with the scale anomaly and the Feymann-Hellmann theorem. The final free parameter left is obtained from quarkonium hadronic transitions~\cite{Pineda:2019mhw}. In Figs.~\ref{lshpp} and \ref{lshkk} we plot the normalized differential widths for the transitions we have computed. The total widths can be found in Eqs.~\eqref{s3:e1}-\eqref{s3:e8}. As in the LO transitions the uncertainty is dominated by multipole expansion corrections, however, unlike the LO case, the gluonic matrix elements have also important uncertainties stemming from the determination of the low-energy constants of the chiral representation.

For the transitions $\Upsilon(11020)\to \Upsilon(nS)\pi^+\pi^-$, $n=1,2,3$, we can compare to the experimental results from Ref.~\cite{Abdesselam:2019gth}. We find remarkable agreement for $n=1$, however the experimental values are larger for $n=2,3$. Nevertheless, the experimental determinations are yet not very precise and future determinations might be closer to our values. We note that the width for $n=1$ is the one with the most precise experimental determination.

In Sec.~\ref{s4} we provide several ratios of transition widths in which some of the uncertainties cancel out. These can be used to test the different approximations made in this paper in an independent way. For instance, the ratios in Eqs.~\eqref{s4a:e1}-\eqref{s4a:e2}, \eqref{s4a:e3} and \eqref{s4a:e4} are independent of the heavy-quark matrix elements but still relying on the factorization of the gluonic matrix elements. Hence, these rations can be used to test for this factorization. Finally, in Sec.~\ref{s5}, we have computed the semi-inclusive widths, generated by the same operators in the Lagrangian in Eq.~\eqref{pnrqcd2} we have considered so far, for the transitions with large enough energy gaps to allow it~\cite{Oncala:2017hop}. The comparison with our results in Secs.~\ref{s2} and \ref{s3} allows us to evaluate the relative importance of the specific light-quark final states for which we have computed transition widths.

Another remarkable result is that the sum of the LO and NLO semi-inclusive widths of $\Upsilon(11020)$ is compatible with the experimental total width. This, combined with the good agreement in the $\Upsilon(11020)\to \Upsilon(1S)\pi^+\pi^-$ transition width and the earlier prediction in Ref.~\cite{Berwein:2015vca} of the mass of a hybrid bottomonium state within $20$~MeV of the current average for the $\Upsilon(11020)$ mass form, in our opinion, strong evidence for the hybrid bottomonium nature of this state.

\section*{Acknowledgements}

J.T.C. acknowledges partial financial support from the European Union's Horizon 2020 research and innovation program under the Marie Sk\l{}odowska--Curie Grant Agreement No. 665919. He has also been supported in part by the Spanish Grants No. FPA2017-86989-P and No. SEV-2016-0588 from the Ministerio de Ciencia, Innovaci\'on y Universidades, and the Grant No. 2017-SGR-1069 from the Generalitat de Catalunya. The work of E.P.\ is supported in part by the U.S. Department of Energy (Contract No. DE-AC05-06OR23177) and National Science Foundation (PHY-1714253).  This research was supported by the Munich Institute for Astro- and Particle Physics (MIAPP) which is funded by the Deutsche Forschungsgemeinschaft (DFG, German Research Foundation) under Germany's Excellence Strategy – EXC-2094 – 390783311.

\appendix

\section{Pseudoscalar production via the axial anomaly}\label{Ap:krollm}

We need the matrix elements of $\bm{E}\cdot\bm{B}$ between the vacuum and $\pi^0$, $\eta$ and $\eta'$. First we note that
\begin{align}
&\frac{g^2}{\pi}\bm{E}\cdot\bm{B}=\alpha_s G_{\mu\nu}\tilde{G}^{\mu\nu}\label{Ap:krollm:e1}\,,
\end{align}
with the dual field-strength tensor defined as $\tilde{G}^{\mu\nu}=\frac{1}{2}\epsilon^{\mu\nu\alpha\beta}G_{\alpha\beta}$ and $\epsilon_{0123}=1$. The matrix elements of $G_{\mu\nu}\tilde{G}^{\mu\nu}$ can then be related to the divergence of the axial current and the pseudoscalar current through the axial anomaly
\begin{align}
\partial^{\mu}J^a_{5\,\mu}=2m_a \bar{a}i\gamma_5a+\frac{\alpha_s}{4\pi}G_{\mu\nu}\tilde{G}^{\mu\nu}\,,\quad a=u,d,s\,,\label{axanom}
\end{align}
with $J^a_{5\,\mu}=\bar{a}\gamma_{\mu}\gamma_5a$. This leaves us with $18$ nonperturbative parameters corresponding to the matrix elements of the axial and pseudoscalar currents $a=u,d,s$ and final states $\pi^0, \eta, \eta'$. This amount of free parameters can be greatly reduced by the implementation of a mixing scheme between $\pi^0-\eta-\eta'$. In the following we review the mixing scheme from Refs.~\cite{Feldmann:1998vh,Kroll:2005sd} that we have used in our computation.

First, let us introduce the notation $|\eta_a\rangle=|a\bar{a}\rangle$, then our primary nonperturbative parameters are $f_a$ and $m^2_{aa}$ defined as follows
\begin{align}
&\langle 0|J^a_{5\,\mu}|\eta_{a'}(p)\rangle=ip_{\mu}f_a\delta_{aa'}\,,\\
&2m_a\langle 0|\bar{a}i\gamma_5a|\eta_{a'}(p)\rangle=f_am^2_{aa}\delta_{aa'}\,,\quad a,a'=u,d,s\,.
\end{align}
We also define the short-hand notation
\begin{align}
\omega_c=\langle 0|\frac{\alpha_s}{4\pi}G_{\mu\nu}\tilde{G}^{\mu\nu}|\eta_{c}(p)\rangle\,,\quad c=\pi^0,\eta,\eta'\label{Ap:krollm:e2}\,.
\end{align}
Since the mixing of the $\pi^0$ with the $\eta$ and $\eta'$ is weak while the $\eta$-$\eta'$ mixing is strong, it is convenient to use isoscalar and isovector combinations of $\eta_u$ and $\eta_d$, that is a change in the basis of states given by
\begin{align}
&\eta_b=M\eta_a,\quad a=u,d,s,\quad b=-,+,s\,,\\
&M=\frac{1}{\sqrt{2}}\left(\begin{array}{ccc} 1 & -1 & 0 \\ 1 & 1 & 0 \\ 0 & 0 & \sqrt{2} \\ \end{array}\right)\,,
\end{align}
where $M$ is an orthogonal matrix. We will consider the physical  $\pi^0$, $\eta$, $\eta'$ states as different mixes of the $\eta_b$, $b=-,+,s$. The unitary matrix that transforms between these two bases is given by
\begin{align}
&\eta_c=U_{cb}\eta_b,\quad b=-,+,s\quad c=\pi^0,\eta,\eta'\,,\\
&U=\left(\begin{array}{ccc} 1 & \beta+\psi\cos\phi & -\psi\sin\phi \\ -\psi-\beta\cos\phi & \cos\phi & -\sin\phi \\ -\beta\sin\phi & \sin\phi & \cos\phi \\ \end{array}\right)\,.\label{krolUM}
\end{align}
Since the mixing of $\pi^0$ with $\eta$ and $\eta'$ is weak, the mixing parameters $\beta$ and $\psi$ are small. Neglecting quadratic terms in these one finds $UU^{\dagger}=1$. Therefore, we can write
\begin{align}
\langle 0|J^b_{5\,\mu}|\eta_{c}(p)\rangle&=M_{ba}ip_{\mu}f_a\delta_{aa'}(M^{\top})_{a'b'}(U^{\dagger})_{b'c}=ip_{\mu}{\cal F}_{bb'}(U^{\dagger})_{b'c}\,,
\end{align}
with 
\begin{align}
{\cal F}&=\left(\begin{array}{ccc} 1 & z & 0 \\ z & 1 & 0 \\ 0 & 0 & 1/y \\ \end{array}\right)\,,\quad z=\frac{f_u-f_d}{f_u+f_d}\,,\quad y=\frac{f_+}{f_s}\,,\quad f_+=\frac{f_u+f_d}{2}\,.
\end{align}
Therefore
\begin{align}
\langle 0|\partial^{\mu}J^b_{5\,\mu}|\eta_{c}(p)\rangle&={\cal F}_{bb'}(U^{\dagger})_{b'c'}{\cal M}_{c'c}={\cal M}_{cc'}U_{c'b'}{\cal F}_{b'b}\,,\label{acs1}
\end{align}
with the mass matrix
\begin{align}
{\cal M}&=\left(\begin{array}{ccc} m^2_{\pi^0} & 0 & 0 \\ 0 & m^2_{\eta} & 0 \\ 0 & 0 & m^2_{\eta'}\end{array}\right)\,.
\end{align}
On the other hand using the axial anomaly from Eq.~\eqref{axanom} we obtain
\begin{align}
\langle 0|\partial^{\mu}J^b_{5\,\mu}|\eta_{c}(p)\rangle&=M_{ba}\left(f_am_{aa}\delta_{aa'}+\omega_{a'}\right) (M^{\top})_{a'b'}(U^{\dagger})_{b'c}=A_{bb'}(U^{\dagger})_{b'c}\,,\label{acs2}
\end{align}
with
\begin{align}
A_{bb'}&=\left(\begin{array}{ccc} \frac{f_um_{uu}+f_dm_{dd}}{2} & \frac{f_um_{uu}-f_dm_{dd}}{2} & 0 \\ \frac{f_um_{uu}-f_dm_{dd}}{2}+\sqrt{2}\omega_- & \frac{f_um_{uu}+f_dm_{dd}}{2}+\sqrt{2}\omega_+ & \sqrt{2}\omega_s \\ \omega_- & \omega_+ & f_sm_{ss}+\omega_s \\ \end{array}\right)\,.
\end{align}
Putting together Eqs.~\eqref{acs1} and \eqref{acs2}
\begin{align}
AU^{\dagger}={\cal F}U^{\dagger}{\cal M}\,,\label{acs3}
\end{align}
we arrive at a system of eight independent equations that we choose to use to determine the mixing parameters $\beta$, $\psi$, $\sin\phi$, the matrix elements $\omega_-$, $\omega_+$, $\omega_s$ and $f_s$ as well as $m_{uu}+m_{dd}$. We obtain
\begin{align}
&m^2_{\pi^0}=\frac{1}{2}\left(m^2_{uu}+m^2_{dd}\right)\,,\label{krolr1}\\
&y=\sqrt{2}\frac{\omega_s}{\omega_+}\,,\\
&z=-\frac{\omega_-}{\omega_+}\,,\\
&\omega_+=f_+\frac{\left(m^2_{\eta}-m^2_{\pi^0}\right)\left(m^2_{\eta'}-m^2_{\pi^0}\right)}{\sqrt{2}\left(m^2_{ss}-m^2_{\pi^0}\right)}\,,\\
&\sin\phi=\sqrt{\frac{\left(m^2_{\eta'}-m^2_{ss}\right)\left(m^2_{\eta}-m^2_{\pi^0}\right)}{\left(m^2_{\eta'}-m^2_{\eta}\right)\left(m^2_{ss}-m^2_{\pi^0}\right)}}\,,\label{sinphi}\\
&y=\sqrt{2\frac{\left(m^2_{\eta'}-m^2_{ss}\right)\left(m^2_{ss}-m^2_{\eta}\right)}{\left(m^2_{\eta'}-m^2_{\pi^0}\right)\left(m^2_{\eta}-m^2_{\pi^0}\right)}}\,,\\
&\beta=z+\frac{m^2_{dd}-m^2_{uu}}{2\left(m^2_{\eta'}-m^2_{\pi^0}\right)}\,,\\
&\psi=\frac{1}{2}\cos\phi\frac{\left(m^2_{dd}-m^2_{uu}\right)\left(m^2_{\eta'}-m^2_{\eta}\right)}{\left(m^2_{\eta'}-m^2_{\pi^0}\right)\left(m^2_{\eta}-m^2_{\pi^0}\right)}\,.\label{krolr8}
\end{align}
Furthermore, $f_+$ is fixed to the pion decay constant $F_{\pi}=92.419$~MeV
\begin{align}
f_+=f_\pi=\sqrt{2}F_{\pi}.
\end{align}
The remaining free parameters are $z$, $m^2_{dd}-m^2_{uu}$ and $m_{ss}$ or $f_s$. As far as we know, the value of $z$ is unknown, however we will not need it as we will see below. The value of $m^2_{dd}-m^2_{uu}$ can be estimated as
\begin{align}
m^2_{dd}-m^2_{uu}=2\left[m^2_{K^0}-m^2_{K^+}-\Delta m^2_{K\,{\rm e.m.}}\right]=0.01248(76)~{\rm GeV}^2\label{mudv}
\end{align}
with the kaon mass difference in QCD taken from Eq.~(9.5) of Ref.~\cite{Colangelo:2018jxw}. The last parameter remaining is $m^2_{ss}$ which is obtained through
\begin{align}
m^2_{ss}=2m^2_{K^0}-m^2_{\pi^0}=0.477019(26)~{\rm GeV}^2\,.
\end{align}

Finally we can obtain the matrix elements $\omega_c$, $c=\pi^0,\eta,\eta'$ using the matrix in Eq.~\eqref{krolUM}, and the relations in Eqs.~\eqref{krolr1}-\eqref{krolr8}:
\begin{align}
\omega_c=U_{cb}\omega_b\,.
\end{align}
We obtain 
\begin{align}
\omega_{\pi^0}&=F_{\pi}\frac{m^2_{dd}-m^2_{uu}}{2}\,,\label{ompi0}\\
\omega_{\eta}&=F_{\pi}(m^2_{\eta}-m^2_{\pi^0})\sqrt{\frac{\left(m^2_{\eta'}-m^2_{\pi^0}\right)\left(m^2_{ss}-m^2_{\eta}\right)}{\left(m^2_{\eta'}-m^2_{\eta}\right)\left(m^2_{ss}-m^2_{\pi^0}\right)}}\,,\label{ometa}\\
\omega_{\eta'}&=F_{\pi}(m^2_{\eta'}-m^2_{\pi^0})\sqrt{\frac{\left(m^2_{\eta}-m^2_{\pi^0}\right)\left(m^2_{\eta'}-m^2_{ss}\right)}{\left(m^2_{\eta'}-m^2_{\eta}\right)\left(m^2_{ss}-m^2_{\pi^0}\right)}}\label{ometap}\,.
\end{align}

Using the meson mass values from the PDG we obtain the following numerical values:
\begin{align}
\omega_{\pi^0}&=0.574(\pm 0.035)\times 10^{-3}~{\rm GeV}^3\,,\label{ompi0num}\\
\omega_{\eta}&=19.5(\pm 0.7)\times 10^{-3}~{\rm GeV}^3\,,\label{ometanum}\\
\omega_{\eta'}&=55(\pm 2)\times 10^{-3}~{\rm GeV}^3\,.\label{ometapnum}
\end{align}
The uncertainty of $\omega_{\pi^0}$ is dominated by the uncertainty of the $m^2_{uu}-m^2_{dd}$ mass difference in Eq.~\eqref{mudv}. The error in not taking into account quadratic terms in $\beta$ and $\psi$ is proportional to $(m^2_{uu}-m^2_{dd})^2$ and negligible in front of the uncertainty of $m^2_{uu}-m^2_{dd}$ itself. In the case of $\omega_{\eta}$ and $\omega_{\eta'}$ the parametric uncertainty is small. Nevertheless, it does not account for the difference between the theoretical mixing angle obtained from Eq.~\eqref{sinphi} $\phi=41.462(4)$º and the phenomenological determination $\phi=39.3$º~\cite{Feldmann:1998vh}. This is likely the result of the model depended approximations of the mixing scheme, as for instance the truncation of the Fock space expansion of $|\eta_a\rangle$ to just the quark-antiquark component. Therefore, we find it more adequate to assign as uncertainty of $\omega_{\eta}$, $\omega_{\eta'}$ the propagation of the error in the determination of the mixing angle $\phi$.

\section{Dispersive representation for two meson production matrix elements}\label{Ap:om}

We want to determine the matrix elements for two pion and two kaon production by $\bm{B}^2$:
\begin{align}
\langle P^+(p_+)P^-(p_-)|g^2\bm{B}^2|0\rangle\,,\quad P=\pi,\,K \label{Ap:om:e1}
\end{align}
First, we note that $\bm{B}^2$ contains both an $S$ and $D$-wave terms. This is more apparent rewriting it as follows:
\begin{align}
\bm{B}^2=\frac{1}{4}G^{\alpha\beta a}G^a_{\alpha\beta}+v^{\mu}v^{\nu}\theta^g_{\mu\nu}\,,\label{Ap:om:e2}
\end{align}
with $v_{\mu}=(1,\,\bm{0})$ and 
\begin{align}
\theta^g_{\mu\nu}=\frac{1}{4}g_{\mu\nu}G^{\alpha\beta a}G^a_{\alpha\beta}-G^a_{\mu\alpha}G^{\alpha\,a}_{\nu}\,.
\end{align}
One can write a chiral representation of the matrix elements in Eq.~\eqref{Ap:om:e1} which will depend on a set of unknown low-energy constants. In order to partially determine these it is useful to rewrite $\bm{B}^2$ as the sum of several terms and write a chiral representation of the matrix elements of each one. The first term in Eq.~\eqref{Ap:om:e2} can be written in terms of the trace of QCD energy-momentum tensor, 
\begin{align}
\bm{B}^2&=\frac{\alpha_s}{\beta(\alpha_s)}\left(\theta^{\mu}_{\mu}-\sum_i m_i(1-\gamma_i)\bar{q}_iq_i\right)+v^{\mu}v^{\nu}\theta^g_{\mu\nu}\,,\label{Ap:om:e3}
\end{align}
with $\gamma_i$ the anomalous dimension of the $\bar{q}_iq_i$ operator, $\beta$ the QCD $\beta$ function and
\begin{align}
\theta^{\mu}_{\mu}=\frac{1}{4}\frac{\beta(\alpha_s)}{\alpha_s}G^{\mu\nu a}G_{\mu\nu a}+\sum_i(1-\gamma_i)m_i\bar{q}_iq_i\,.\label{Ap:om:e13}
\end{align}
Now, one can write chiral representations for the matrix elements of each one of the terms in Eq.~\eqref{Ap:om:e3}. At LO in the chiral expansion each one of these matrix elements depends on only one low-energy constant (the normalization of the matrix element). Furthermore, for the $\theta^{\mu}_{\mu}$ and $m_i\bar{q}_iq_i$ these can be determined by the scale anomaly~\cite{Voloshin:1980zf,Novikov:1980fa,Chivukula:1989ds} and the Feymann-Hellmann theorem, respectively. At LO these matrix elements read as
\begin{align}
\langle P^+(p_+)P^-(p_-)|\theta^{\mu}_{\mu}|0\rangle&=2(p_+\cdot p_-)+4m^2_{P}+\dots\quad P=\pi,\,K,\\
\langle P^+(p_+)P^-(p_-)|\sum_i m_i\bar{q}_iq_i|0\rangle&=m^2_{P}+\dots,\\
\langle P^+(p_+)P^-(p_-)|\theta^{g}_{\mu\nu}|0\rangle&=-V_2(\mu)\left(p_{+\mu}p_{-\nu}+p_{-\mu}p_{+\nu}-\frac{1}{2}g_{\mu\nu}p_+\cdot p_-+\dots\right)\,.
\end{align}
Adding up all the contributions and neglecting the anomalous dimension and contributions to the $\beta$ function beyond the LO we arrive at
\begin{align}
\langle P^+(p_+)P^-(p_-)|\frac{\beta_0\alpha_s}{2\pi}\bm{B}^2|0\rangle&=-\left[\left(2-\frac{3\kappa}{2}\right)p_+\cdot p_-+6\kappa p^0_+p^0_-+3m^2_{P}\right]\,,\label{Ap:om:e4}
\end{align}
where we have used the definition $\kappa=\alpha_s \beta_0 V_2(\mu)/(6\pi)$ as in Ref.~\cite{Novikov:1980fa}. The parameter $\kappa$ cannot be determined from first principles, however it can be extracted from the spectrum of the transitions $\psi(2S)\to J/\psi\,\pi^+\pi^-$ and $\Upsilon(2S)\to \Upsilon(1S)\,\pi^+\pi^-$~\cite{Novikov:1980fa,Pineda:2019mhw}. We use the value from Ref.~\cite{Pineda:2019mhw}
\begin{align}
\kappa=0.247(20)\,.\label{Ap:om:e12}
\end{align}

Using Eqs.~\eqref{epem}, \eqref{Ap:mvw:e6} and \eqref{Ap:mvw:e7} we can write Eq.~\eqref{Ap:om:e4} in terms of the Mandelstam variables:
\begin{align}
\langle P^+(p_+)P^-(p_-)|\frac{\beta_0\alpha_s}{2\pi}\bm{B}^2|0\rangle&=-\left[\left(1-\frac{3\kappa}{4}\right)s+\left(1+\frac{3}{2}\kappa\right)m^2_{P}+\frac{3\kappa}{2}\left(\Delta^2-\left(\frac{u-t}{2m_n}\right)^2\right)\right]\,,\label{Ap:om:e9}
\end{align}
where $\Delta=m_n-m_m$  and $m_n$ and $m_m$ are the masses of the initial hybrid and the final standard quarkonium states, respectively.

The chiral representation we have just built is only valid at low energies $\sqrt{s}\ll\Lambda_{\chi}$. For the computation of the transitions we need these matrix elements up to $\sqrt{s}\simeq 1.6$~GeV. To do so we will build a dispersive representation of the matrix elements. Let us define the form factors
\begin{align}
F_P\equiv \langle P^+(p_+)P^-(p_-)|\frac{\beta_0\alpha_s}{2\pi}\bm{B}^2|0\rangle,\,\quad P=\pi,K\,.\label{Ap:om:e11}
\end{align}
We can write a general decomposition of these form factors, in the spirit of the reconstruction theorem~\cite{Stern:1993rg}, considering their analytic properties. In our case this is greatly simplified since only the cuts in the right-hand side of the complex $s$-plane corresponding to two-pion and two-kaon rescattering need to be considered. Furthermore, we know that the form factors only contain $S$- and $D$-waves. Hence, a general decomposition of the form factors is as follows
\begin{align}
F_P(s,t,u)=F^{(0)}_P(s)+\left[(u-t)^2-\frac{4}{3}m^2_n\sigma_P^2(s)\rho_P^2(s)\right]F^{(2)}_P(s)\,,\label{Ap:om:e10}
\end{align}
with $\sigma_P$ and $\rho_P$ defined in Eqs.~\eqref{Ap:mvw:e8} and \eqref{rrep2}, respectively. From Watson's theorem~\cite{Watson:1954uc} the discontinuity along the cut of the form factors $F^{(l)}_P$ generated by the two-pion or kaon rescattering has the following form
\begin{align}
&\text{Im}\left[n_PF^{(l)}_P(s)\right]=\sum_{P'=\pi,K}(T^{0*}_l(s))_{PP'}\sigma_{P'}(s)n_{P'}F^{(l)}_{P'}(s)\theta(s-4m^2_{P'})\,,\label{Ap:om:e5}
\end{align}
where $n_\pi=\sqrt{3/2}$ and $n_K=\sqrt{2}$ are factors resulting from the projection of the pion and kaon states into isospin $I=0$. The T-matrix $T^0_l(s)$ is given by
\begin{align}
\bm{T}^0_l(s)=\left(\begin{array}{cc}\frac{\eta^0_l(s)e^{2i\delta^0_l(s)}-1}{2i\sigma_{\pi}(s)} & |g^0_l(s)|e^{i\psi^0_l(s)} \\ |g^0_l(s)|e^{i\psi^0_l(s)} & \frac{\eta^0_l(s)e^{2i(\psi^0_l(s)-\delta^0_l(s))}-1}{2i\sigma_{K}(s)}\end{array} \right)\,.\label{Ap:om:e5b}
\end{align}
The three inputs of the T-matrix are: the $l$-wave isoscalar $\pi\pi$ phase shift $\delta^0_l(s)$ and the modulus, $|g^0_l|$, and phase, $\psi^0_l(s)$, of the $l$-wave isoscalar $\pi\pi\to K\bar{K}$ amplitude. The inelasticity $\eta^0_l(s)$ is related to $|g^0_l|$ by
\begin{align}
\eta^0_l(s)=\sqrt{1-4|g^0_l(s)|^2\sigma_{\pi}(s)\sigma_K(s)\theta(s-4m^2_K)}\,.
\end{align}
We want to find a functional form of the form factors that fulfills Eq.~\eqref{Ap:om:e5}, is analytic in the complex $s$-plane, except on the cuts, and is real on the real $s$ axis below the cuts. This is the two-channel Muskhelishvili-Omn\`es problem~\cite{mushi,Omnes:1958hv}. There are two independent canonical solutions~\cite{mushi,Donoghue:1990xh} which we arrange as columns of the following matrix
\begin{align}
\bm{\Omega}^{(l)}(s)=\left(\begin{array}{cc} C^{(l)}_{1}(s) & D^{(l)}_{1}(s) \\ C^{(l)}_{2}(s) & D^{(l)}_{2}(s) \end{array}\right)\,.
\end{align}
A general solution can be written as
\begin{align}
n_P F^{(l)}_P(s)=\Omega^{(l)}_{PP'}(s)Q^{(l)}_{P'}(s)\,,\label{Ap:om:e8}
\end{align}
where $\bm{Q}^{(l)}(s)=(Q^{(l)}_1,\,Q^{(l)}_2)$ are the so-called subtraction polynomials. The $\Omega$-matrix satisfies a set of coupled Muskhelishvili-Omn\`es singular integral equations
\begin{align}
\bm{\Omega}(s)=\frac{1}{\pi}\int^{\infty}_{4m^2_{\pi}}\frac{ds'}{s'-s}\left(\bm{T}^0_l(s')\right)^{*}\bm{\Sigma}(s')\bm{\Omega}(s')\,,\label{Ap:om:e6}
\end{align}
with $\bm{\Sigma}(s)=\text{diag}(\sigma_\pi(s)\theta(s-4m^2_{\pi}),\sigma_K(s)\theta(s-4m^2_{K}))$. The two independent solutions are generated choosing the normalization $\bm{\Omega}(0)=\mathbb{1}$.

In the limit $s\to\infty$ we expect the form factors to go to zero as $1/s$, therefore $\bm{\Omega}^{(l)}\bm{Q}^{(l)}$ should also vanish in the same way. If in this limit $\bm{\Omega}^{(l)}\sim 1/s^r$, then $\bm{Q}^{(l)}$ should be a degree $(r-1)$ polynomial. If we take the determinant of both sides of Eq.~\eqref{Ap:om:e6} the matrix equation reduces to a one-dimensional equation for which an analytical solution is available~\cite{mushi,Omnes:1958hv}. The asymptotic behavior can then be obtained~\cite{Moussallam:1999aq}:
\begin{align}
\text{det}(\bm{\Omega}^{(l)})\stackrel{s\to\infty}{\sim}s^{-\text{Arg}(\text{det}(S^0_l))/\pi} \label{Ap:om:e7}
\end{align}
with $S^0_l$ the S-matrix associated to the $T$-matrix in Eq.~\eqref{Ap:om:e5b}. Assuming that the off-diagonal terms of $S^0_l$ vanish in the asymptotic limit, then $\text{Arg}(\text{det}(S^0_l))$ is just the sum of the asymptotic behaviors of the eigen phase shifts. Since each component of $\bm{\Omega}$ must vanish at least as $1/s$, Eq.~\eqref{Ap:om:e7} establishes a constraint on the asymptotic behavior of the $T$-matrix in order for solutions of the Muskhelishvili-Omn\`es integral equations to exist
\begin{align}
\lim_{s\to\infty}\text{Arg}(\text{det}(S^0_l(s)))\geq n\pi\,,
\end{align}
with $n$ the number of open channels considered. In the present work we only consider the two-pion and two-kaon channels (i.e. $n=2$), which is an approximation valid only up to a certain value of $s$, therefore any given asymptotic behavior of the T-matrix can only be considered as a model, which we choose to ensure the existence of solutions of Eq.~\eqref{Ap:om:e6}.

The inputs of the $T$-matrix are taken as follows: $\delta^0_l(s)$ is taken from the parametrization of Ref.~\cite{GarciaMartin:2011cn} with the CFD parameter set; $|g^0_l|$ and $\psi^0_l(s)$ are taken as the parametrizations from Ref.~\cite{Pelaez:2018qny} with the CFD$_\text{c}$ and CFD parameter sets for $S$ and $D$ waves, respectively. These parametrizations are given up to $\sqrt{s}=1.42$~GeV for $\delta^0_l(s)$ and $\sqrt{s}=2$~GeV for $|g^0_l|$ and $\psi^0_l(s)$, then they are continued smoothly, up to first derivatives, to the following asymptotic values $\delta^0_l\to 2\pi$, $\psi^0_l\to 2\pi$ and $|g^0_l|\to 0$.

\begin{figure}[ht!]
\begin{tabular}{cc}
\includegraphics[width=.48\textwidth]{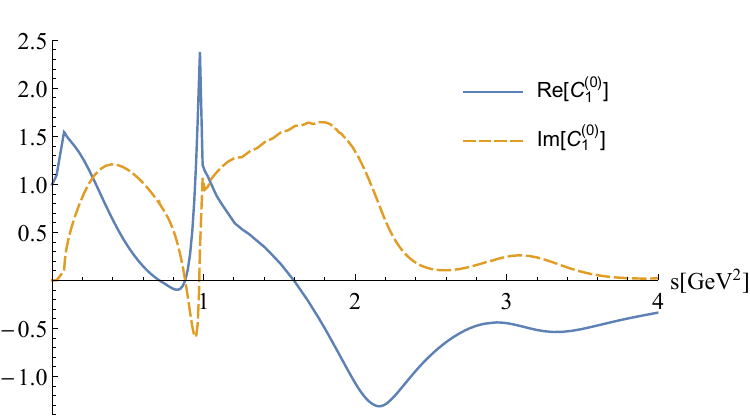} & \includegraphics[width=.48\textwidth]{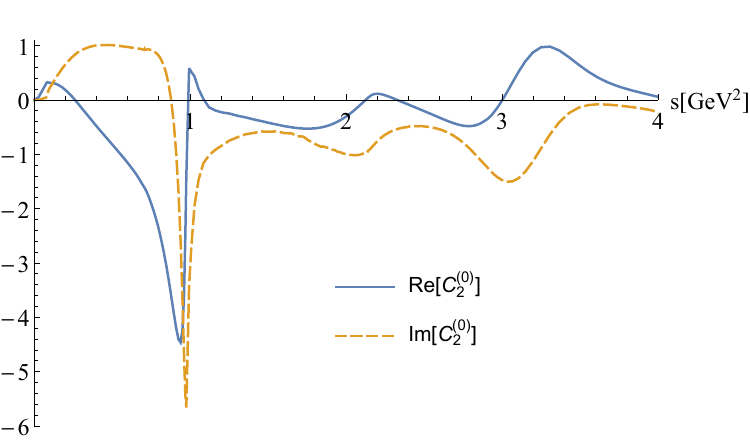}   \\
\includegraphics[width=.48\textwidth]{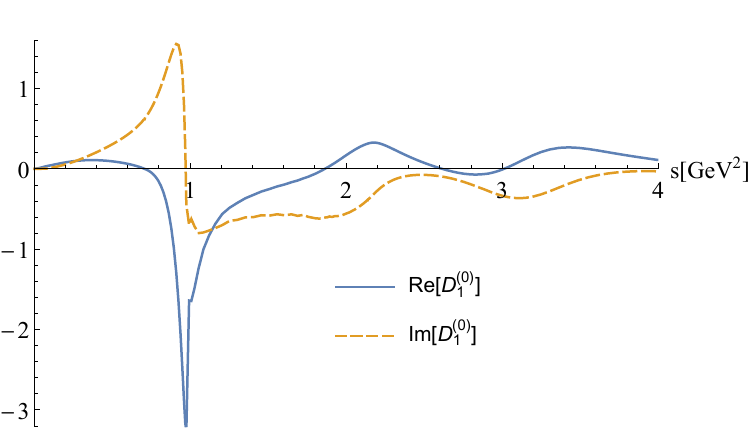} & \includegraphics[width=.48\textwidth]{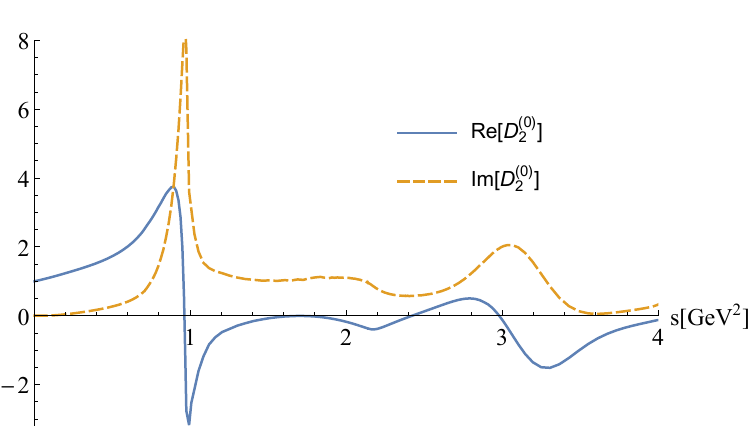}   \\
\end{tabular}
\caption{Plot of the solutions $\bm{\Omega}^{(0)}$ of the two-pion and two-kaon coupled Muskhelishvili-Omn\`es equations for the $S$ partial wave.}
\label{omatrixs}
\begin{tabular}{cc}
\includegraphics[width=.48\textwidth]{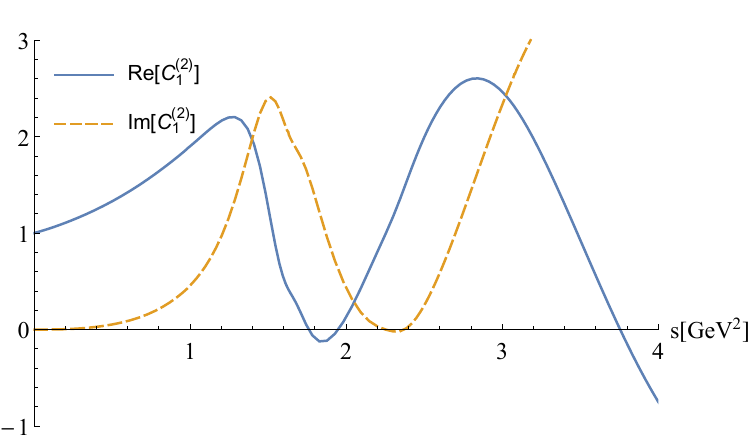} & \includegraphics[width=.48\textwidth]{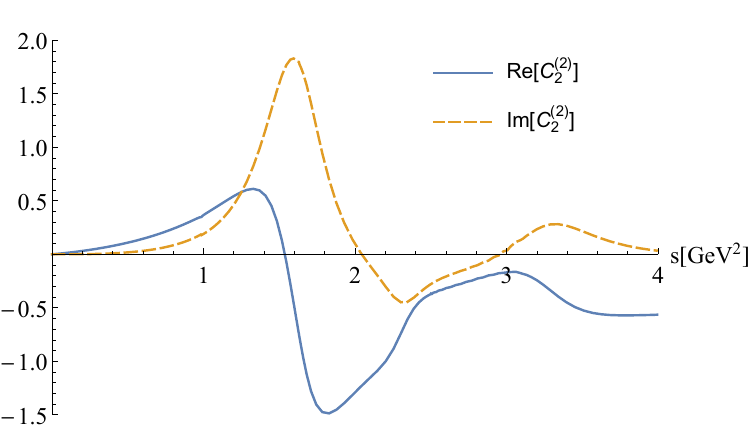}   \\
\includegraphics[width=.48\textwidth]{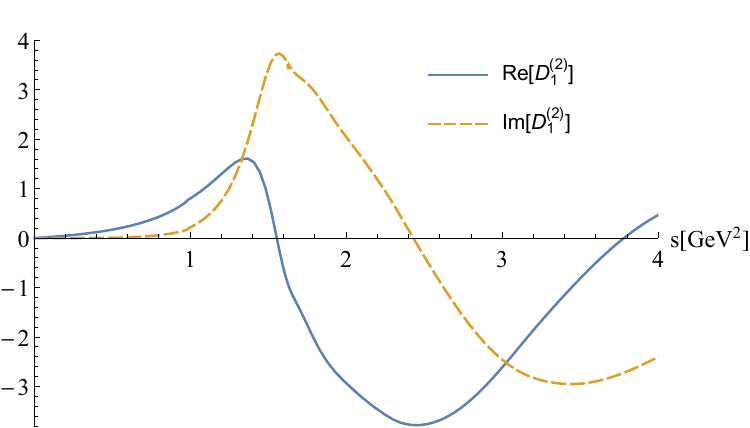} & \includegraphics[width=.48\textwidth]{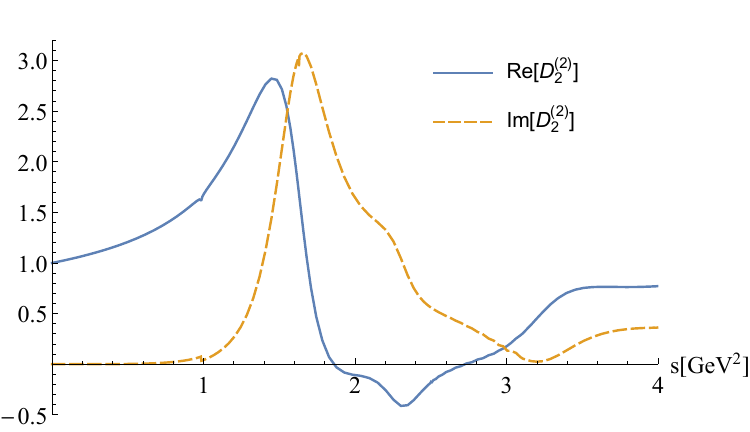}   \\
\end{tabular}
\caption{Plot of the solutions $\bm{\Omega}^{(2)}$ of the two-pion and two-kaon coupled Muskhelishvili-Omn\`es equations for the $D$ partial wave.}
	\label{omatrixd}
\end{figure}

The solution of Eq.~\eqref{Ap:om:e6} is obtained numerically using the procedure described in Refs.~\cite{Moussallam:1999aq,Descotes-Genon:2000pfd}. A very brief summary is as follows.\footnote{A detailed explanation can be found in Ref.~\cite{Descotes-Genon:2000pfd}.} First, one rewrites Eq.~\eqref{Ap:om:e6} in terms of $\text{Re}(\bm{\Omega})$ only. Then the dispersive integral is split in $j=1,\dots,M$ subintervals and  the numerator of the integrand is expanded in Legendre polynomials up to degree $N$. This allows the exact evaluation of the principal value integration in terms of Legendre functions of the second kind. The coefficients of the Legendre expansions for each subinterval $j$ are integrated using $z^{(j)}_i$, $i=1,\dots,N$ Gauss-Legendre points. This determines $\text{Re}[\bm{\Omega}(s)]$ in terms of $\text{Re}[\bm{\Omega}(z^{(j)}_i)]$. Evaluating $\text{Re}[\bm{\Omega}(s)]$ precisely at the same Gauss-Legendre points generates a system of equations with $\text{Re}[\bm{\Omega}(z^{(j)}_i)]$ as variables. Adding the normalization, an overdetermined system $(NM+2)\times NM$ is created. Using a singular value decomposition one can obtain a pseudoinverse of the matrix and the variables are obtained in a least squares fit. The imaginary parts can then be obtained from the unitarity condition in Eq.~\eqref{Ap:om:e5}. We use $N=25$, eight subintervals for the $S$-wave case and six for the $D$-wave one. We plot our results in Figs.~\ref{omatrixs} and ~\ref{omatrixd}. For the $S$-wave case we have checked that our results agree with Refs.~\cite{Moussallam:1999aq,Celis:2013xja}. The $D$-wave results are presented here for the first time.

The final step in the construction of the dispersive representation of the matrix elements in Eq.~\eqref{Ap:om:e1} is the determination of the $\bm{Q}^{(l)}(s)$ polynomials. We do so by requiring that the expansion of Eq.~\eqref{Ap:om:e8} for small $s$ matches the chiral representation of Eq.~\eqref{Ap:om:e9}. In principle, given the asymptotic behavior of the $T$-matrix we have constructed, one would expect that these polynomials are just constants. However, this does not produce appropriate results~\cite{Donoghue:1990xh} as the LO chiral representation in Eq.~\eqref{Ap:om:e9} depends on $s$ and cannot be reproduced by the dispersive representation unless one allows for $\bm{Q}^{(l)}(s)$ polynomials of order one. Doing so spoils the asymptotic behavior of the dispersive representation of the form factors. However, since we are only interested in these form factors up to $\sqrt{s}\sim 1.6$~GeV the issue can be ignored. Furthermore, due to the presence of a $D$ wave, the partial wave projected chiral amplitude in Eq.~\eqref{Ap:om:e9} contains a singular $1/s$ term. To accommodate it an analogous $1/s$ needs to be added to the $\bm{Q}^{(0)}$ polynomials~\cite{Chen:2015jgl}. The resulting $\bm{Q}^{(l)}$ are as follows
\begin{align}
Q^{(0)}_1&=a^{(0)}_{11}s^{-1}+a^{(0)}_{12}+a^{(0)}_{13}s\,,\\
Q^{(0)}_2&=a^{(0)}_{21}s^{-1}+a^{(0)}_{22}+a^{(0)}_{23}s\,,\\
Q^{(2)}_1&=a^{(2)}_{1}\,,\\
Q^{(2)}_2&=a^{(2)}_{2}\,,
\end{align}
and the coefficients
\begin{align}
a^{(0)}_{11}=&-\sqrt{6}m^2_{\pi}\kappa\Delta^2\,,\label{Ap:om:sp:e1}\\
a^{(0)}_{12}=&-\sqrt{\frac{3}{2}}\left[m^2_{\pi}\left(1-\frac{\kappa}{2}\right)+\kappa\,\Delta^2\left(1-2m_{\pi}^2\dot{C}_1(0)-\frac{4}{\sqrt{3}}m^2_K\dot{D}_1(0)\right)\right]\,,\\
a^{(0)}_{13}=&-\sqrt{\frac{3}{2}}\left\{1-m^2_{\pi}\dot{C}_1(0)-\frac{2}{\sqrt{3}}m^2_K\dot{D}_1(0)-\frac{\kappa}{4}\left[1-2m^2_{\pi}\dot{C}_1(0)-\frac{4}{\sqrt{3}}m^2_K\dot{D}_1(0)\right]-\kappa\,\Delta^2\left[\dot{C}_1(0)+\frac{2}{\sqrt{3}}\dot{D}_1(0)\right.\right.\nn\\
&\left.\left.+m^2_{\pi}\left(\ddot{C}_1(0)-2\left(\dot{C}^2_1(0)+\dot{C}_2(0)\dot{D}_1(0)\right)\right)+\frac{2}{\sqrt{3}}m^2_K\left(\ddot{D}_1(0)-2\dot{D}_1(0)\left(\dot{C}_1(0)+\dot{D}_2(0)\right)\right)\right]\right\}\,,\\
a^{(0)}_{21}=&-2\sqrt{2}m^2_K\kappa\Delta^2\,,\\
a^{(0)}_{22}=&-\sqrt{2}\left[m^2_{K}\left(1-\frac{\kappa}{2}\right)+\kappa\,\Delta^2\left(1-\sqrt{3}m_{\pi}^2\dot{C}_2(0)-2m^2_K\dot{D}_2(0)\right)\right]\,,\\
a^{(0)}_{23}=&-\sqrt{2}\left\{1-\frac{\sqrt{3}}{2}m^2_{\pi}\dot{C}_2(0)-m^2_K\dot{D}_2(0)-\frac{\kappa}{4}\left[1-\sqrt{3}m^2_{\pi}\dot{C}_2(0)-2m^2_K\dot{D}_2(0)\right]-\kappa\,\Delta^2\left[\frac{\sqrt{3}}{2}\dot{C}_2(0)+\dot{D}_2(0)\right.\right.\nn\\
&\left.\left.+\frac{\sqrt{3}}{2}m^2_{\pi}\left(\ddot{C}_2(0)-2\dot{C}_2(0)\left(\dot{C}_1(0)+\dot{D}_2(0)\right)\right)+m^2_K\left(\ddot{D}_2(0)-2\left(\dot{C}_2(0)\dot{D}_1(0)+\dot{D}^2_2(0)\right)\right)\right]\right\}\,,\\
a^{(2)}_{1}=&\sqrt{\frac{3}{2}}\frac{3\kappa}{8m^2_n}\,,\\
a^{(2)}_{2}=&\sqrt{2}\frac{3\kappa}{8m^2_n}\,.\label{Ap:om:sp:e2}
\end{align}

\section{Mandelstam variables and width formulas}\label{Ap:mvw}

The Mandelstam variables for the transition $H_{n}(k_n)\to S_{m}(k_m)P^+(p_+)P^-(p_-)$ with $P=\pi,\,K$ are as follows
\begin{align}
s=(p_++p_-)^2,\,\quad t=(k_n-p_+)^2,\,\quad u=(k_n-p_-)^2\,.\label{Ap:mvw:e1}
\end{align}
In the reference frame of the decaying exotic quarkonia, one finds 
\begin{align}
p_\pm^0=\frac{1}{2}\left(\Delta\pm\rho(s)\sigma_P(s)\cos\theta\right)\,,\label{epem} 
\end{align}
with 
\begin{align}
\sigma_P&=\sqrt{1-4m_{P}^2/s}\,,\label{Ap:mvw:e8}\\
\Delta&=\frac{m^2_{n}-m^2_{m}+s}{2m_{n}} \,,\label{rrep1}\\
\rho&=\sqrt{\Delta^2-s}\label{rrep2}\,,
\end{align}
where $m_n$ and $m_{m}$ are the masses of initial and final quarkonium respectively. In the nonrelativistic approximation of the final quarkonium momentum, the above expressions for $\Delta$ reduce to
\begin{align}
&\Delta=m_{n}-m_{m} \,.\label{rrep3}
\end{align}
In all our numerical computations we have used this nonrelativistic approximation of $\Delta$ to be consistent with the nonrelativistic nature of our EFT approach. It should be noted that the specific form of Eqs.~\eqref{Ap:om:sp:e1}-\eqref{Ap:om:sp:e2} depends on this choice.

Continuing in the reference frame of the decaying exotic quarkonia and using Eq.~\eqref{epem} one finds
\begin{align}
t=&\frac{1}{2}\left(m^2_n-m^2_{m}+2m^2_P-s\right)-m_n\rho(s)\sigma_P(s)\cos\theta\,,\label{Ap:mvw:e6}\\
u=&\frac{1}{2}\left(m^2_n-m^2_{m}+2m^2_P-s\right)+m_n\rho(s)\sigma_P(s)\cos\theta\,,\label{Ap:mvw:e7}
\end{align}
and consequently
\begin{align}
\cos\theta=\frac{u-t}{2m_n\rho(s)\sigma_P(s)}\,.\label{Ap:mvw:e2}
\end{align}

We are interested in finding the transition width of amplitudes with an $S$ and $D$ wave as in Eq.~\eqref{Ap:om:e10}
\begin{align}
\mathcal{A}_{2P}&=\mathcal{A}^{(0)}(s)+\left[(u-t)^2-\frac{4}{3}m^2_n\sigma_P^2(s)\rho_P^2(s)\right]\mathcal{A}^{(2)}(s)\,.\label{Ap:mvw:e3}
\end{align}
Using Eq.~\eqref{Ap:mvw:e2} into Eq.~\eqref{Ap:mvw:e3} we can write $\mathcal{A}_{2P}=\mathcal{A}_{2P}(s,\cos\theta)$.

The differential decay width is
\begin{align}
\frac{\Gamma_{2P}}{d s\,d\cos\theta}&=\frac{\rho(s)\sigma_P(s) }{8(2\pi)^3}\left|\mathcal{A}_{2P}(s,\cos\theta)\right|^2\,.
\end{align}
Integrating $\theta$ we arrive at
\begin{align}
\frac{d\Gamma_{2P}}{ds}&=\frac{\rho(s)\sigma_P(s)}{8(2\pi)^3}\left[2\left|\mathcal{A}^{(0)}(s)\right|^2+\frac{8}{45}(2m_n\sigma_P(s)\rho(s))^2\left|\mathcal{A}^{(2)}(s)\right|^2\right]\,.\label{Ap:mvw:e4}
\end{align}

To obtain the total decay width we integrate numerically 
\begin{align}
\Gamma_{2P}=\int^{(m_{n}-m_{m})^2}_{4m^2_{P}}ds\left(\frac{d\Gamma_{2P}}{ds}\right)\,.\label{Ap:mvw:e5}
\end{align}

In the transition with one light-quark meson in the final state, the momenta are fixed by momentum conservation. The final light-quark meson momentum is $|\bm{p}_P|=\rho(m_P)$, and the decay width is given by
\begin{align}
\Gamma_P=\frac{\rho(m^2_{P})}{2\pi}|\mathcal{A}_{1P}|^2\,,\label{dw1p}
\end{align}
with $\rho(s)$ given in Eq.~\eqref{rrep2}.

\bibliographystyle{apsrev4-1}
\bibliography{hybtrans}

\begin{thebibliography}{57}%
\makeatletter
\providecommand \@ifxundefined [1]{%
 \@ifx{#1\undefined}
}%
\providecommand \@ifnum [1]{%
 \ifnum #1\expandafter \@firstoftwo
 \else \expandafter \@secondoftwo
 \fi
}%
\providecommand \@ifx [1]{%
 \ifx #1\expandafter \@firstoftwo
 \else \expandafter \@secondoftwo
 \fi
}%
\providecommand \natexlab [1]{#1}%
\providecommand \enquote  [1]{``#1''}%
\providecommand \bibnamefont  [1]{#1}%
\providecommand \bibfnamefont [1]{#1}%
\providecommand \citenamefont [1]{#1}%
\providecommand \href@noop [0]{\@secondoftwo}%
\providecommand \href [0]{\begingroup \@sanitize@url \@href}%
\providecommand \@href[1]{\@@startlink{#1}\@@href}%
\providecommand \@@href[1]{\endgroup#1\@@endlink}%
\providecommand \@sanitize@url [0]{\catcode `\\12\catcode `\$12\catcode
  `\&12\catcode `\#12\catcode `\^12\catcode `\_12\catcode `\%12\relax}%
\providecommand \@@startlink[1]{}%
\providecommand \@@endlink[0]{}%
\providecommand \url  [0]{\begingroup\@sanitize@url \@url }%
\providecommand \@url [1]{\endgroup\@href {#1}{\urlprefix }}%
\providecommand \urlprefix  [0]{URL }%
\providecommand \Eprint [0]{\href }%
\providecommand \doibase [0]{http://dx.doi.org/}%
\providecommand \selectlanguage [0]{\@gobble}%
\providecommand \bibinfo  [0]{\@secondoftwo}%
\providecommand \bibfield  [0]{\@secondoftwo}%
\providecommand \translation [1]{[#1]}%
\providecommand \BibitemOpen [0]{}%
\providecommand \bibitemStop [0]{}%
\providecommand \bibitemNoStop [0]{.\EOS\space}%
\providecommand \EOS [0]{\spacefactor3000\relax}%
\providecommand \BibitemShut  [1]{\csname bibitem#1\endcsname}%
\let\auto@bib@innerbib\@empty
%</preamble>
\bibitem [{\citenamefont {Gell-Mann}(1964)}]{GellMann:1964nj}%
  \BibitemOpen
  \bibfield  {author} {\bibinfo {author} {\bibfnamefont {M.}~\bibnamefont
  {Gell-Mann}},\ }\href {\doibase 10.1016/S0031-9163(64)92001-3} {\bibfield
  {journal} {\bibinfo  {journal} {Phys. Lett.}\ }\textbf {\bibinfo {volume}
  {8}},\ \bibinfo {pages} {214} (\bibinfo {year} {1964})}\BibitemShut {NoStop}%
\bibitem [{\citenamefont {Choi}\ \emph {et~al.}(2003)\citenamefont {Choi} \emph
  {et~al.}}]{Choi:2003ue}%
  \BibitemOpen
  \bibfield  {author} {\bibinfo {author} {\bibfnamefont {S.~K.}\ \bibnamefont
  {Choi}} \emph {et~al.} (\bibinfo {collaboration} {Belle}),\ }\href {\doibase
  10.1103/PhysRevLett.91.262001} {\bibfield  {journal} {\bibinfo  {journal}
  {Phys. Rev. Lett.}\ }\textbf {\bibinfo {volume} {91}},\ \bibinfo {pages}
  {262001} (\bibinfo {year} {2003})},\ \Eprint
  {http://arxiv.org/abs/hep-ex/0309032} {arXiv:hep-ex/0309032} \BibitemShut
  {NoStop}%
\bibitem [{\citenamefont {Caswell}\ and\ \citenamefont
  {Lepage}(1986)}]{Caswell:1985ui}%
  \BibitemOpen
  \bibfield  {author} {\bibinfo {author} {\bibfnamefont {W.~E.}\ \bibnamefont
  {Caswell}}\ and\ \bibinfo {author} {\bibfnamefont {G.~P.}\ \bibnamefont
  {Lepage}},\ }\href {\doibase 10.1016/0370-2693(86)91297-9} {\bibfield
  {journal} {\bibinfo  {journal} {Phys. Lett. B}\ }\textbf {\bibinfo {volume}
  {167}},\ \bibinfo {pages} {437} (\bibinfo {year} {1986})}\BibitemShut
  {NoStop}%
\bibitem [{\citenamefont {Bodwin}\ \emph {et~al.}(1995)\citenamefont {Bodwin},
  \citenamefont {Braaten},\ and\ \citenamefont {Lepage}}]{Bodwin:1994jh}%
  \BibitemOpen
  \bibfield  {author} {\bibinfo {author} {\bibfnamefont {G.~T.}\ \bibnamefont
  {Bodwin}}, \bibinfo {author} {\bibfnamefont {E.}~\bibnamefont {Braaten}}, \
  and\ \bibinfo {author} {\bibfnamefont {G.~P.}\ \bibnamefont {Lepage}},\
  }\href {\doibase 10.1103/PhysRevD.55.5853} {\bibfield  {journal} {\bibinfo
  {journal} {Phys. Rev. D}\ }\textbf {\bibinfo {volume} {51}},\ \bibinfo
  {pages} {1125} (\bibinfo {year} {1995})},\ \bibinfo {note} {[Erratum:
  Phys.Rev.D 55, 5853 (1997)]},\ \Eprint {http://arxiv.org/abs/hep-ph/9407339}
  {arXiv:hep-ph/9407339} \BibitemShut {NoStop}%
\bibitem [{\citenamefont {Berwein}\ \emph {et~al.}(2015)\citenamefont
  {Berwein}, \citenamefont {Brambilla}, \citenamefont {Tarr\'us~Castell\`a},\
  and\ \citenamefont {Vairo}}]{Berwein:2015vca}%
  \BibitemOpen
  \bibfield  {author} {\bibinfo {author} {\bibfnamefont {M.}~\bibnamefont
  {Berwein}}, \bibinfo {author} {\bibfnamefont {N.}~\bibnamefont {Brambilla}},
  \bibinfo {author} {\bibfnamefont {J.}~\bibnamefont {Tarr\'us~Castell\`a}}, \
  and\ \bibinfo {author} {\bibfnamefont {A.}~\bibnamefont {Vairo}},\ }\href
  {\doibase 10.1103/PhysRevD.92.114019} {\bibfield  {journal} {\bibinfo
  {journal} {Phys. Rev. D}\ }\textbf {\bibinfo {volume} {92}},\ \bibinfo
  {pages} {114019} (\bibinfo {year} {2015})},\ \Eprint
  {http://arxiv.org/abs/1510.04299} {arXiv:1510.04299 [hep-ph]} \BibitemShut
  {NoStop}%
\bibitem [{\citenamefont {Juge}\ \emph {et~al.}(2003)\citenamefont {Juge},
  \citenamefont {Kuti},\ and\ \citenamefont {Morningstar}}]{Juge:2002br}%
  \BibitemOpen
  \bibfield  {author} {\bibinfo {author} {\bibfnamefont {K.~J.}\ \bibnamefont
  {Juge}}, \bibinfo {author} {\bibfnamefont {J.}~\bibnamefont {Kuti}}, \ and\
  \bibinfo {author} {\bibfnamefont {C.}~\bibnamefont {Morningstar}},\ }\href
  {\doibase 10.1103/PhysRevLett.90.161601} {\bibfield  {journal} {\bibinfo
  {journal} {Phys. Rev. Lett.}\ }\textbf {\bibinfo {volume} {90}},\ \bibinfo
  {pages} {161601} (\bibinfo {year} {2003})},\ \Eprint
  {http://arxiv.org/abs/hep-lat/0207004} {arXiv:hep-lat/0207004} \BibitemShut
  {NoStop}%
\bibitem [{\citenamefont {Capitani}\ \emph {et~al.}(2019)\citenamefont
  {Capitani}, \citenamefont {Philipsen}, \citenamefont {Reisinger},
  \citenamefont {Riehl},\ and\ \citenamefont {Wagner}}]{Capitani:2018rox}%
  \BibitemOpen
  \bibfield  {author} {\bibinfo {author} {\bibfnamefont {S.}~\bibnamefont
  {Capitani}}, \bibinfo {author} {\bibfnamefont {O.}~\bibnamefont {Philipsen}},
  \bibinfo {author} {\bibfnamefont {C.}~\bibnamefont {Reisinger}}, \bibinfo
  {author} {\bibfnamefont {C.}~\bibnamefont {Riehl}}, \ and\ \bibinfo {author}
  {\bibfnamefont {M.}~\bibnamefont {Wagner}},\ }\href {\doibase
  10.1103/PhysRevD.99.034502} {\bibfield  {journal} {\bibinfo  {journal} {Phys.
  Rev. D}\ }\textbf {\bibinfo {volume} {99}},\ \bibinfo {pages} {034502}
  (\bibinfo {year} {2019})},\ \Eprint {http://arxiv.org/abs/1811.11046}
  {arXiv:1811.11046 [hep-lat]} \BibitemShut {NoStop}%
\bibitem [{\citenamefont {Brambilla}\ \emph {et~al.}(2018)\citenamefont
  {Brambilla}, \citenamefont {Krein}, \citenamefont {Tarr\'us~Castell\`a},\
  and\ \citenamefont {Vairo}}]{Brambilla:2017uyf}%
  \BibitemOpen
  \bibfield  {author} {\bibinfo {author} {\bibfnamefont {N.}~\bibnamefont
  {Brambilla}}, \bibinfo {author} {\bibfnamefont {G.}~\bibnamefont {Krein}},
  \bibinfo {author} {\bibfnamefont {J.}~\bibnamefont {Tarr\'us~Castell\`a}}, \
  and\ \bibinfo {author} {\bibfnamefont {A.}~\bibnamefont {Vairo}},\ }\href
  {\doibase 10.1103/PhysRevD.97.016016} {\bibfield  {journal} {\bibinfo
  {journal} {Phys. Rev. D}\ }\textbf {\bibinfo {volume} {97}},\ \bibinfo
  {pages} {016016} (\bibinfo {year} {2018})},\ \Eprint
  {http://arxiv.org/abs/1707.09647} {arXiv:1707.09647 [hep-ph]} \BibitemShut
  {NoStop}%
\bibitem [{\citenamefont {Oncala}\ and\ \citenamefont
  {Soto}(2017)}]{Oncala:2017hop}%
  \BibitemOpen
  \bibfield  {author} {\bibinfo {author} {\bibfnamefont {R.}~\bibnamefont
  {Oncala}}\ and\ \bibinfo {author} {\bibfnamefont {J.}~\bibnamefont {Soto}},\
  }\href {\doibase 10.1103/PhysRevD.96.014004} {\bibfield  {journal} {\bibinfo
  {journal} {Phys. Rev. D}\ }\textbf {\bibinfo {volume} {96}},\ \bibinfo
  {pages} {014004} (\bibinfo {year} {2017})},\ \Eprint
  {http://arxiv.org/abs/1702.03900} {arXiv:1702.03900 [hep-ph]} \BibitemShut
  {NoStop}%
\bibitem [{\citenamefont {Brambilla}\ \emph {et~al.}(2019)\citenamefont
  {Brambilla}, \citenamefont {Lai}, \citenamefont {Segovia}, \citenamefont
  {Tarr\'us~Castell\`a},\ and\ \citenamefont {Vairo}}]{Brambilla:2018pyn}%
  \BibitemOpen
  \bibfield  {author} {\bibinfo {author} {\bibfnamefont {N.}~\bibnamefont
  {Brambilla}}, \bibinfo {author} {\bibfnamefont {W.~K.}\ \bibnamefont {Lai}},
  \bibinfo {author} {\bibfnamefont {J.}~\bibnamefont {Segovia}}, \bibinfo
  {author} {\bibfnamefont {J.}~\bibnamefont {Tarr\'us~Castell\`a}}, \ and\
  \bibinfo {author} {\bibfnamefont {A.}~\bibnamefont {Vairo}},\ }\href
  {\doibase 10.1103/PhysRevD.99.014017} {\bibfield  {journal} {\bibinfo
  {journal} {Phys. Rev. D}\ }\textbf {\bibinfo {volume} {99}},\ \bibinfo
  {pages} {014017} (\bibinfo {year} {2019})},\ \bibinfo {note} {[Erratum:
  Phys.Rev.D 101, 099902 (2020)]},\ \Eprint {http://arxiv.org/abs/1805.07713}
  {arXiv:1805.07713 [hep-ph]} \BibitemShut {NoStop}%
\bibitem [{\citenamefont {Brambilla}\ \emph {et~al.}(2020)\citenamefont
  {Brambilla}, \citenamefont {Lai}, \citenamefont {Segovia},\ and\
  \citenamefont {Tarr\'us~Castell\`a}}]{Brambilla:2019jfi}%
  \BibitemOpen
  \bibfield  {author} {\bibinfo {author} {\bibfnamefont {N.}~\bibnamefont
  {Brambilla}}, \bibinfo {author} {\bibfnamefont {W.~K.}\ \bibnamefont {Lai}},
  \bibinfo {author} {\bibfnamefont {J.}~\bibnamefont {Segovia}}, \ and\
  \bibinfo {author} {\bibfnamefont {J.}~\bibnamefont {Tarr\'us~Castell\`a}},\
  }\href {\doibase 10.1103/PhysRevD.101.054040} {\bibfield  {journal} {\bibinfo
   {journal} {Phys. Rev. D}\ }\textbf {\bibinfo {volume} {101}},\ \bibinfo
  {pages} {054040} (\bibinfo {year} {2020})},\ \Eprint
  {http://arxiv.org/abs/1908.11699} {arXiv:1908.11699 [hep-ph]} \BibitemShut
  {NoStop}%
\bibitem [{\citenamefont {Griffiths}\ \emph {et~al.}(1983)\citenamefont
  {Griffiths}, \citenamefont {Michael},\ and\ \citenamefont
  {Rakow}}]{Griffiths:1983ah}%
  \BibitemOpen
  \bibfield  {author} {\bibinfo {author} {\bibfnamefont {L.~A.}\ \bibnamefont
  {Griffiths}}, \bibinfo {author} {\bibfnamefont {C.}~\bibnamefont {Michael}},
  \ and\ \bibinfo {author} {\bibfnamefont {P.~E.~L.}\ \bibnamefont {Rakow}},\
  }\href {\doibase 10.1016/0370-2693(83)90680-9} {\bibfield  {journal}
  {\bibinfo  {journal} {Phys. Lett. B}\ }\textbf {\bibinfo {volume} {129}},\
  \bibinfo {pages} {351} (\bibinfo {year} {1983})}\BibitemShut {NoStop}%
\bibitem [{\citenamefont {Juge}\ \emph {et~al.}(1999)\citenamefont {Juge},
  \citenamefont {Kuti},\ and\ \citenamefont {Morningstar}}]{Juge:1999ie}%
  \BibitemOpen
  \bibfield  {author} {\bibinfo {author} {\bibfnamefont {K.~J.}\ \bibnamefont
  {Juge}}, \bibinfo {author} {\bibfnamefont {J.}~\bibnamefont {Kuti}}, \ and\
  \bibinfo {author} {\bibfnamefont {C.~J.}\ \bibnamefont {Morningstar}},\
  }\href {\doibase 10.1103/PhysRevLett.82.4400} {\bibfield  {journal} {\bibinfo
   {journal} {Phys. Rev. Lett.}\ }\textbf {\bibinfo {volume} {82}},\ \bibinfo
  {pages} {4400} (\bibinfo {year} {1999})},\ \Eprint
  {http://arxiv.org/abs/hep-ph/9902336} {arXiv:hep-ph/9902336} \BibitemShut
  {NoStop}%
\bibitem [{\citenamefont {Guo}\ \emph {et~al.}(2008)\citenamefont {Guo},
  \citenamefont {Szczepaniak}, \citenamefont {Galata}, \citenamefont
  {Vassallo},\ and\ \citenamefont {Santopinto}}]{Guo:2008yz}%
  \BibitemOpen
  \bibfield  {author} {\bibinfo {author} {\bibfnamefont {P.}~\bibnamefont
  {Guo}}, \bibinfo {author} {\bibfnamefont {A.~P.}\ \bibnamefont
  {Szczepaniak}}, \bibinfo {author} {\bibfnamefont {G.}~\bibnamefont {Galata}},
  \bibinfo {author} {\bibfnamefont {A.}~\bibnamefont {Vassallo}}, \ and\
  \bibinfo {author} {\bibfnamefont {E.}~\bibnamefont {Santopinto}},\ }\href
  {\doibase 10.1103/PhysRevD.78.056003} {\bibfield  {journal} {\bibinfo
  {journal} {Phys. Rev. D}\ }\textbf {\bibinfo {volume} {78}},\ \bibinfo
  {pages} {056003} (\bibinfo {year} {2008})},\ \Eprint
  {http://arxiv.org/abs/0807.2721} {arXiv:0807.2721 [hep-ph]} \BibitemShut
  {NoStop}%
\bibitem [{\citenamefont {Braaten}\ \emph {et~al.}(2014)\citenamefont
  {Braaten}, \citenamefont {Langmack},\ and\ \citenamefont
  {Smith}}]{Braaten:2014qka}%
  \BibitemOpen
  \bibfield  {author} {\bibinfo {author} {\bibfnamefont {E.}~\bibnamefont
  {Braaten}}, \bibinfo {author} {\bibfnamefont {C.}~\bibnamefont {Langmack}}, \
  and\ \bibinfo {author} {\bibfnamefont {D.~H.}\ \bibnamefont {Smith}},\ }\href
  {\doibase 10.1103/PhysRevD.90.014044} {\bibfield  {journal} {\bibinfo
  {journal} {Phys. Rev. D}\ }\textbf {\bibinfo {volume} {90}},\ \bibinfo
  {pages} {014044} (\bibinfo {year} {2014})},\ \Eprint
  {http://arxiv.org/abs/1402.0438} {arXiv:1402.0438 [hep-ph]} \BibitemShut
  {NoStop}%
\bibitem [{\citenamefont {M\"uller}\ \emph {et~al.}(2019)\citenamefont
  {M\"uller}, \citenamefont {Philipsen}, \citenamefont {Reisinger},\ and\
  \citenamefont {Wagner}}]{Mueller:2019mkh}%
  \BibitemOpen
  \bibfield  {author} {\bibinfo {author} {\bibfnamefont {L.}~\bibnamefont
  {M\"uller}}, \bibinfo {author} {\bibfnamefont {O.}~\bibnamefont {Philipsen}},
  \bibinfo {author} {\bibfnamefont {C.}~\bibnamefont {Reisinger}}, \ and\
  \bibinfo {author} {\bibfnamefont {M.}~\bibnamefont {Wagner}},\ }\href
  {\doibase 10.1103/PhysRevD.100.054503} {\bibfield  {journal} {\bibinfo
  {journal} {Phys. Rev. D}\ }\textbf {\bibinfo {volume} {100}},\ \bibinfo
  {pages} {054503} (\bibinfo {year} {2019})},\ \Eprint
  {http://arxiv.org/abs/1907.01482} {arXiv:1907.01482 [hep-lat]} \BibitemShut
  {NoStop}%
\bibitem [{\citenamefont {Soto}\ and\ \citenamefont
  {Tarr\'us~Castell\`a}(2020{\natexlab{a}})}]{Soto:2020xpm}%
  \BibitemOpen
  \bibfield  {author} {\bibinfo {author} {\bibfnamefont {J.}~\bibnamefont
  {Soto}}\ and\ \bibinfo {author} {\bibfnamefont {J.}~\bibnamefont
  {Tarr\'us~Castell\`a}},\ }\href {\doibase 10.1103/PhysRevD.102.014012}
  {\bibfield  {journal} {\bibinfo  {journal} {Phys. Rev. D}\ }\textbf {\bibinfo
  {volume} {102}},\ \bibinfo {pages} {014012} (\bibinfo {year}
  {2020}{\natexlab{a}})},\ \Eprint {http://arxiv.org/abs/2005.00552}
  {arXiv:2005.00552 [hep-ph]} \BibitemShut {NoStop}%
\bibitem [{\citenamefont {Soto}\ and\ \citenamefont
  {Tarr\'us~Castell\`a}(2020{\natexlab{b}})}]{Soto:2020pfa}%
  \BibitemOpen
  \bibfield  {author} {\bibinfo {author} {\bibfnamefont {J.}~\bibnamefont
  {Soto}}\ and\ \bibinfo {author} {\bibfnamefont {J.}~\bibnamefont
  {Tarr\'us~Castell\`a}},\ }\href {\doibase 10.1103/PhysRevD.102.014013}
  {\bibfield  {journal} {\bibinfo  {journal} {Phys. Rev. D}\ }\textbf {\bibinfo
  {volume} {102}},\ \bibinfo {pages} {014013} (\bibinfo {year}
  {2020}{\natexlab{b}})},\ \Eprint {http://arxiv.org/abs/2005.00551}
  {arXiv:2005.00551 [hep-ph]} \BibitemShut {NoStop}%
\bibitem [{\citenamefont {Bali}\ \emph {et~al.}(2000)\citenamefont {Bali},
  \citenamefont {Bolder}, \citenamefont {Eicker}, \citenamefont {Lippert},
  \citenamefont {Orth}, \citenamefont {Ueberholz}, \citenamefont {Schilling},\
  and\ \citenamefont {Struckmann}}]{Bali:2000vr}%
  \BibitemOpen
  \bibfield  {author} {\bibinfo {author} {\bibfnamefont {G.~S.}\ \bibnamefont
  {Bali}}, \bibinfo {author} {\bibfnamefont {B.}~\bibnamefont {Bolder}},
  \bibinfo {author} {\bibfnamefont {N.}~\bibnamefont {Eicker}}, \bibinfo
  {author} {\bibfnamefont {T.}~\bibnamefont {Lippert}}, \bibinfo {author}
  {\bibfnamefont {B.}~\bibnamefont {Orth}}, \bibinfo {author} {\bibfnamefont
  {P.}~\bibnamefont {Ueberholz}}, \bibinfo {author} {\bibfnamefont
  {K.}~\bibnamefont {Schilling}}, \ and\ \bibinfo {author} {\bibfnamefont
  {T.}~\bibnamefont {Struckmann}} (\bibinfo {collaboration} {TXL, T(X)L}),\
  }\href {\doibase 10.1103/PhysRevD.62.054503} {\bibfield  {journal} {\bibinfo
  {journal} {Phys. Rev. D}\ }\textbf {\bibinfo {volume} {62}},\ \bibinfo
  {pages} {054503} (\bibinfo {year} {2000})},\ \Eprint
  {http://arxiv.org/abs/hep-lat/0003012} {arXiv:hep-lat/0003012} \BibitemShut
  {NoStop}%
\bibitem [{\citenamefont {Guo}\ \emph {et~al.}(2018)\citenamefont {Guo},
  \citenamefont {Hanhart}, \citenamefont {Mei\ss{}ner}, \citenamefont {Wang},
  \citenamefont {Zhao},\ and\ \citenamefont {Zou}}]{Guo:2017jvc}%
  \BibitemOpen
  \bibfield  {author} {\bibinfo {author} {\bibfnamefont {F.-K.}\ \bibnamefont
  {Guo}}, \bibinfo {author} {\bibfnamefont {C.}~\bibnamefont {Hanhart}},
  \bibinfo {author} {\bibfnamefont {U.-G.}\ \bibnamefont {Mei\ss{}ner}},
  \bibinfo {author} {\bibfnamefont {Q.}~\bibnamefont {Wang}}, \bibinfo {author}
  {\bibfnamefont {Q.}~\bibnamefont {Zhao}}, \ and\ \bibinfo {author}
  {\bibfnamefont {B.-S.}\ \bibnamefont {Zou}},\ }\href {\doibase
  10.1103/RevModPhys.90.015004} {\bibfield  {journal} {\bibinfo  {journal}
  {Rev. Mod. Phys.}\ }\textbf {\bibinfo {volume} {90}},\ \bibinfo {pages}
  {015004} (\bibinfo {year} {2018})},\ \Eprint
  {http://arxiv.org/abs/1705.00141} {arXiv:1705.00141 [hep-ph]} \BibitemShut
  {NoStop}%
\bibitem [{\citenamefont {Bali}\ \emph {et~al.}(2005)\citenamefont {Bali},
  \citenamefont {Neff}, \citenamefont {Duessel}, \citenamefont {Lippert},\ and\
  \citenamefont {Schilling}}]{Bali:2005fu}%
  \BibitemOpen
  \bibfield  {author} {\bibinfo {author} {\bibfnamefont {G.~S.}\ \bibnamefont
  {Bali}}, \bibinfo {author} {\bibfnamefont {H.}~\bibnamefont {Neff}}, \bibinfo
  {author} {\bibfnamefont {T.}~\bibnamefont {Duessel}}, \bibinfo {author}
  {\bibfnamefont {T.}~\bibnamefont {Lippert}}, \ and\ \bibinfo {author}
  {\bibfnamefont {K.}~\bibnamefont {Schilling}} (\bibinfo {collaboration}
  {SESAM}),\ }\href {\doibase 10.1103/PhysRevD.71.114513} {\bibfield  {journal}
  {\bibinfo  {journal} {Phys. Rev. D}\ }\textbf {\bibinfo {volume} {71}},\
  \bibinfo {pages} {114513} (\bibinfo {year} {2005})},\ \Eprint
  {http://arxiv.org/abs/hep-lat/0505012} {arXiv:hep-lat/0505012} \BibitemShut
  {NoStop}%
\bibitem [{\citenamefont {Bulava}\ \emph {et~al.}(2019)\citenamefont {Bulava},
  \citenamefont {H\"orz}, \citenamefont {Knechtli}, \citenamefont {Koch},
  \citenamefont {Moir}, \citenamefont {Morningstar},\ and\ \citenamefont
  {Peardon}}]{Bulava:2019iut}%
  \BibitemOpen
  \bibfield  {author} {\bibinfo {author} {\bibfnamefont {J.}~\bibnamefont
  {Bulava}}, \bibinfo {author} {\bibfnamefont {B.}~\bibnamefont {H\"orz}},
  \bibinfo {author} {\bibfnamefont {F.}~\bibnamefont {Knechtli}}, \bibinfo
  {author} {\bibfnamefont {V.}~\bibnamefont {Koch}}, \bibinfo {author}
  {\bibfnamefont {G.}~\bibnamefont {Moir}}, \bibinfo {author} {\bibfnamefont
  {C.}~\bibnamefont {Morningstar}}, \ and\ \bibinfo {author} {\bibfnamefont
  {M.}~\bibnamefont {Peardon}},\ }\href {\doibase
  10.1016/j.physletb.2019.05.018} {\bibfield  {journal} {\bibinfo  {journal}
  {Phys. Lett. B}\ }\textbf {\bibinfo {volume} {793}},\ \bibinfo {pages} {493}
  (\bibinfo {year} {2019})},\ \Eprint {http://arxiv.org/abs/1902.04006}
  {arXiv:1902.04006 [hep-lat]} \BibitemShut {NoStop}%
\bibitem [{\citenamefont {Pineda}\ and\ \citenamefont
  {Soto}(1998)}]{Pineda:1997bj}%
  \BibitemOpen
  \bibfield  {author} {\bibinfo {author} {\bibfnamefont {A.}~\bibnamefont
  {Pineda}}\ and\ \bibinfo {author} {\bibfnamefont {J.}~\bibnamefont {Soto}},\
  }\href {\doibase 10.1016/S0920-5632(97)01102-X} {\bibfield  {journal}
  {\bibinfo  {journal} {Nucl. Phys. B Proc. Suppl.}\ }\textbf {\bibinfo
  {volume} {64}},\ \bibinfo {pages} {428} (\bibinfo {year} {1998})},\ \Eprint
  {http://arxiv.org/abs/hep-ph/9707481} {arXiv:hep-ph/9707481} \BibitemShut
  {NoStop}%
\bibitem [{\citenamefont {Brambilla}\ \emph {et~al.}(2000)\citenamefont
  {Brambilla}, \citenamefont {Pineda}, \citenamefont {Soto},\ and\
  \citenamefont {Vairo}}]{Brambilla:1999xf}%
  \BibitemOpen
  \bibfield  {author} {\bibinfo {author} {\bibfnamefont {N.}~\bibnamefont
  {Brambilla}}, \bibinfo {author} {\bibfnamefont {A.}~\bibnamefont {Pineda}},
  \bibinfo {author} {\bibfnamefont {J.}~\bibnamefont {Soto}}, \ and\ \bibinfo
  {author} {\bibfnamefont {A.}~\bibnamefont {Vairo}},\ }\href {\doibase
  10.1016/S0550-3213(99)00693-8} {\bibfield  {journal} {\bibinfo  {journal}
  {Nucl. Phys. B}\ }\textbf {\bibinfo {volume} {566}},\ \bibinfo {pages} {275}
  (\bibinfo {year} {2000})},\ \Eprint {http://arxiv.org/abs/hep-ph/9907240}
  {arXiv:hep-ph/9907240} \BibitemShut {NoStop}%
\bibitem [{\citenamefont {Peset}\ \emph
  {et~al.}(2018{\natexlab{a}})\citenamefont {Peset}, \citenamefont {Pineda},\
  and\ \citenamefont {Segovia}}]{Peset:2018jkf}%
  \BibitemOpen
  \bibfield  {author} {\bibinfo {author} {\bibfnamefont {C.}~\bibnamefont
  {Peset}}, \bibinfo {author} {\bibfnamefont {A.}~\bibnamefont {Pineda}}, \
  and\ \bibinfo {author} {\bibfnamefont {J.}~\bibnamefont {Segovia}},\ }\href
  {\doibase 10.1103/PhysRevD.98.094003} {\bibfield  {journal} {\bibinfo
  {journal} {Phys. Rev. D}\ }\textbf {\bibinfo {volume} {98}},\ \bibinfo
  {pages} {094003} (\bibinfo {year} {2018}{\natexlab{a}})},\ \Eprint
  {http://arxiv.org/abs/1809.09124} {arXiv:1809.09124 [hep-ph]} \BibitemShut
  {NoStop}%
\bibitem [{\citenamefont {Santel}\ \emph {et~al.}(2016)\citenamefont {Santel}
  \emph {et~al.}}]{Santel:2015qga}%
  \BibitemOpen
  \bibfield  {author} {\bibinfo {author} {\bibfnamefont {D.}~\bibnamefont
  {Santel}} \emph {et~al.} (\bibinfo {collaboration} {Belle}),\ }\href
  {\doibase 10.1103/PhysRevD.93.011101} {\bibfield  {journal} {\bibinfo
  {journal} {Phys. Rev. D}\ }\textbf {\bibinfo {volume} {93}},\ \bibinfo
  {pages} {011101} (\bibinfo {year} {2016})},\ \Eprint
  {http://arxiv.org/abs/1501.01137} {arXiv:1501.01137 [hep-ex]} \BibitemShut
  {NoStop}%
\bibitem [{\citenamefont {Abdesselam}\ \emph {et~al.}(2016)\citenamefont
  {Abdesselam} \emph {et~al.}}]{Abdesselam:2015zza}%
  \BibitemOpen
  \bibfield  {author} {\bibinfo {author} {\bibfnamefont {A.}~\bibnamefont
  {Abdesselam}} \emph {et~al.} (\bibinfo {collaboration} {Belle}),\ }\href
  {\doibase 10.1103/PhysRevLett.117.142001} {\bibfield  {journal} {\bibinfo
  {journal} {Phys. Rev. Lett.}\ }\textbf {\bibinfo {volume} {117}},\ \bibinfo
  {pages} {142001} (\bibinfo {year} {2016})},\ \Eprint
  {http://arxiv.org/abs/1508.06562} {arXiv:1508.06562 [hep-ex]} \BibitemShut
  {NoStop}%
\bibitem [{\citenamefont {Mizuk}\ \emph {et~al.}(2019)\citenamefont {Mizuk}
  \emph {et~al.}}]{Abdesselam:2019gth}%
  \BibitemOpen
  \bibfield  {author} {\bibinfo {author} {\bibfnamefont {R.}~\bibnamefont
  {Mizuk}} \emph {et~al.} (\bibinfo {collaboration} {Belle}),\ }\href {\doibase
  10.1007/JHEP10(2019)220} {\bibfield  {journal} {\bibinfo  {journal} {JHEP}\
  }\textbf {\bibinfo {volume} {10}},\ \bibinfo {pages} {220} (\bibinfo {year}
  {2019})},\ \Eprint {http://arxiv.org/abs/1905.05521} {arXiv:1905.05521
  [hep-ex]} \BibitemShut {NoStop}%
\bibitem [{\citenamefont {Bondar}\ \emph {et~al.}(2012)\citenamefont {Bondar}
  \emph {et~al.}}]{Belle:2011aa}%
  \BibitemOpen
  \bibfield  {author} {\bibinfo {author} {\bibfnamefont {A.}~\bibnamefont
  {Bondar}} \emph {et~al.} (\bibinfo {collaboration} {Belle}),\ }\href
  {\doibase 10.1103/PhysRevLett.108.122001} {\bibfield  {journal} {\bibinfo
  {journal} {Phys. Rev. Lett.}\ }\textbf {\bibinfo {volume} {108}},\ \bibinfo
  {pages} {122001} (\bibinfo {year} {2012})},\ \Eprint
  {http://arxiv.org/abs/1110.2251} {arXiv:1110.2251 [hep-ex]} \BibitemShut
  {NoStop}%
\bibitem [{\citenamefont {Pineda}\ and\ \citenamefont
  {Tarr\'us~Castell\`a}(2019)}]{Pineda:2019mhw}%
  \BibitemOpen
  \bibfield  {author} {\bibinfo {author} {\bibfnamefont {A.}~\bibnamefont
  {Pineda}}\ and\ \bibinfo {author} {\bibfnamefont {J.}~\bibnamefont
  {Tarr\'us~Castell\`a}},\ }\href {\doibase 10.1103/PhysRevD.100.054021}
  {\bibfield  {journal} {\bibinfo  {journal} {Phys. Rev. D}\ }\textbf {\bibinfo
  {volume} {100}},\ \bibinfo {pages} {054021} (\bibinfo {year} {2019})},\
  \Eprint {http://arxiv.org/abs/1905.03794} {arXiv:1905.03794 [hep-ph]}
  \BibitemShut {NoStop}%
\bibitem [{\citenamefont {Feldmann}\ \emph {et~al.}(1998)\citenamefont
  {Feldmann}, \citenamefont {Kroll},\ and\ \citenamefont
  {Stech}}]{Feldmann:1998vh}%
  \BibitemOpen
  \bibfield  {author} {\bibinfo {author} {\bibfnamefont {T.}~\bibnamefont
  {Feldmann}}, \bibinfo {author} {\bibfnamefont {P.}~\bibnamefont {Kroll}}, \
  and\ \bibinfo {author} {\bibfnamefont {B.}~\bibnamefont {Stech}},\ }\href
  {\doibase 10.1103/PhysRevD.58.114006} {\bibfield  {journal} {\bibinfo
  {journal} {Phys. Rev. D}\ }\textbf {\bibinfo {volume} {58}},\ \bibinfo
  {pages} {114006} (\bibinfo {year} {1998})},\ \Eprint
  {http://arxiv.org/abs/hep-ph/9802409} {arXiv:hep-ph/9802409} \BibitemShut
  {NoStop}%
\bibitem [{\citenamefont {Kroll}(2005)}]{Kroll:2005sd}%
  \BibitemOpen
  \bibfield  {author} {\bibinfo {author} {\bibfnamefont {P.}~\bibnamefont
  {Kroll}},\ }\href {\doibase 10.1142/S0217732305018633} {\bibfield  {journal}
  {\bibinfo  {journal} {Mod. Phys. Lett. A}\ }\textbf {\bibinfo {volume}
  {20}},\ \bibinfo {pages} {2667} (\bibinfo {year} {2005})},\ \Eprint
  {http://arxiv.org/abs/hep-ph/0509031} {arXiv:hep-ph/0509031} \BibitemShut
  {NoStop}%
\bibitem [{\citenamefont {Donoghue}\ \emph {et~al.}(1990)\citenamefont
  {Donoghue}, \citenamefont {Gasser},\ and\ \citenamefont
  {Leutwyler}}]{Donoghue:1990xh}%
  \BibitemOpen
  \bibfield  {author} {\bibinfo {author} {\bibfnamefont {J.~F.}\ \bibnamefont
  {Donoghue}}, \bibinfo {author} {\bibfnamefont {J.}~\bibnamefont {Gasser}}, \
  and\ \bibinfo {author} {\bibfnamefont {H.}~\bibnamefont {Leutwyler}},\ }\href
  {\doibase 10.1016/0550-3213(90)90474-R} {\bibfield  {journal} {\bibinfo
  {journal} {Nucl. Phys. B}\ }\textbf {\bibinfo {volume} {343}},\ \bibinfo
  {pages} {341} (\bibinfo {year} {1990})}\BibitemShut {NoStop}%
\bibitem [{\citenamefont {Moussallam}(2000)}]{Moussallam:1999aq}%
  \BibitemOpen
  \bibfield  {author} {\bibinfo {author} {\bibfnamefont {B.}~\bibnamefont
  {Moussallam}},\ }\href {\doibase 10.1007/s100520050738} {\bibfield  {journal}
  {\bibinfo  {journal} {Eur. Phys. J. C}\ }\textbf {\bibinfo {volume} {14}},\
  \bibinfo {pages} {111} (\bibinfo {year} {2000})},\ \Eprint
  {http://arxiv.org/abs/hep-ph/9909292} {arXiv:hep-ph/9909292} \BibitemShut
  {NoStop}%
\bibitem [{\citenamefont {Celis}\ \emph {et~al.}(2014)\citenamefont {Celis},
  \citenamefont {Cirigliano},\ and\ \citenamefont {Passemar}}]{Celis:2013xja}%
  \BibitemOpen
  \bibfield  {author} {\bibinfo {author} {\bibfnamefont {A.}~\bibnamefont
  {Celis}}, \bibinfo {author} {\bibfnamefont {V.}~\bibnamefont {Cirigliano}}, \
  and\ \bibinfo {author} {\bibfnamefont {E.}~\bibnamefont {Passemar}},\ }\href
  {\doibase 10.1103/PhysRevD.89.013008} {\bibfield  {journal} {\bibinfo
  {journal} {Phys. Rev. D}\ }\textbf {\bibinfo {volume} {89}},\ \bibinfo
  {pages} {013008} (\bibinfo {year} {2014})},\ \Eprint
  {http://arxiv.org/abs/1309.3564} {arXiv:1309.3564 [hep-ph]} \BibitemShut
  {NoStop}%
\bibitem [{\citenamefont {Foster}\ and\ \citenamefont
  {Michael}(1999)}]{Foster:1998wu}%
  \BibitemOpen
  \bibfield  {author} {\bibinfo {author} {\bibfnamefont {M.}~\bibnamefont
  {Foster}}\ and\ \bibinfo {author} {\bibfnamefont {C.}~\bibnamefont {Michael}}
  (\bibinfo {collaboration} {UKQCD}),\ }\href {\doibase
  10.1103/PhysRevD.59.094509} {\bibfield  {journal} {\bibinfo  {journal} {Phys.
  Rev. D}\ }\textbf {\bibinfo {volume} {59}},\ \bibinfo {pages} {094509}
  (\bibinfo {year} {1999})},\ \Eprint {http://arxiv.org/abs/hep-lat/9811010}
  {arXiv:hep-lat/9811010} \BibitemShut {NoStop}%
\bibitem [{\citenamefont {Bali}\ and\ \citenamefont
  {Pineda}(2004)}]{Bali:2003jq}%
  \BibitemOpen
  \bibfield  {author} {\bibinfo {author} {\bibfnamefont {G.~S.}\ \bibnamefont
  {Bali}}\ and\ \bibinfo {author} {\bibfnamefont {A.}~\bibnamefont {Pineda}},\
  }\href {\doibase 10.1103/PhysRevD.69.094001} {\bibfield  {journal} {\bibinfo
  {journal} {Phys. Rev. D}\ }\textbf {\bibinfo {volume} {69}},\ \bibinfo
  {pages} {094001} (\bibinfo {year} {2004})},\ \Eprint
  {http://arxiv.org/abs/hep-ph/0310130} {arXiv:hep-ph/0310130} \BibitemShut
  {NoStop}%
\bibitem [{\citenamefont {Ayala}\ \emph {et~al.}(2020)\citenamefont {Ayala},
  \citenamefont {Lobregat},\ and\ \citenamefont {Pineda}}]{Ayala:2020pxq}%
  \BibitemOpen
  \bibfield  {author} {\bibinfo {author} {\bibfnamefont {C.}~\bibnamefont
  {Ayala}}, \bibinfo {author} {\bibfnamefont {X.}~\bibnamefont {Lobregat}}, \
  and\ \bibinfo {author} {\bibfnamefont {A.}~\bibnamefont {Pineda}},\ }\href
  {\doibase 10.1007/JHEP12(2020)093} {\bibfield  {journal} {\bibinfo  {journal}
  {JHEP}\ }\textbf {\bibinfo {volume} {12}},\ \bibinfo {pages} {093} (\bibinfo
  {year} {2020})},\ \Eprint {http://arxiv.org/abs/2009.01285} {arXiv:2009.01285
  [hep-ph]} \BibitemShut {NoStop}%
\bibitem [{\citenamefont {Bramon}\ \emph {et~al.}(1999)\citenamefont {Bramon},
  \citenamefont {Escribano},\ and\ \citenamefont {Scadron}}]{Bramon:1997va}%
  \BibitemOpen
  \bibfield  {author} {\bibinfo {author} {\bibfnamefont {A.}~\bibnamefont
  {Bramon}}, \bibinfo {author} {\bibfnamefont {R.}~\bibnamefont {Escribano}}, \
  and\ \bibinfo {author} {\bibfnamefont {M.~D.}\ \bibnamefont {Scadron}},\
  }\href {\doibase 10.1007/s100529801009} {\bibfield  {journal} {\bibinfo
  {journal} {Eur. Phys. J. C}\ }\textbf {\bibinfo {volume} {7}},\ \bibinfo
  {pages} {271} (\bibinfo {year} {1999})},\ \Eprint
  {http://arxiv.org/abs/hep-ph/9711229} {arXiv:hep-ph/9711229} \BibitemShut
  {NoStop}%
\bibitem [{\citenamefont {Escribano}\ and\ \citenamefont
  {Royo}(2020)}]{Escribano:2020jdy}%
  \BibitemOpen
  \bibfield  {author} {\bibinfo {author} {\bibfnamefont {R.}~\bibnamefont
  {Escribano}}\ and\ \bibinfo {author} {\bibfnamefont {E.}~\bibnamefont
  {Royo}},\ }\href {\doibase 10.1016/j.physletb.2020.135534} {\bibfield
  {journal} {\bibinfo  {journal} {Phys. Lett. B}\ }\textbf {\bibinfo {volume}
  {807}},\ \bibinfo {pages} {135534} (\bibinfo {year} {2020})},\ \Eprint
  {http://arxiv.org/abs/2003.08379} {arXiv:2003.08379 [hep-ph]} \BibitemShut
  {NoStop}%
\bibitem [{\citenamefont {Chen}\ \emph {et~al.}(2017)\citenamefont {Chen},
  \citenamefont {Cleven}, \citenamefont {Daub}, \citenamefont {Guo},
  \citenamefont {Hanhart}, \citenamefont {Kubis}, \citenamefont {Mei\ss{}ner},\
  and\ \citenamefont {Zou}}]{Chen:2016mjn}%
  \BibitemOpen
  \bibfield  {author} {\bibinfo {author} {\bibfnamefont {Y.-H.}\ \bibnamefont
  {Chen}}, \bibinfo {author} {\bibfnamefont {M.}~\bibnamefont {Cleven}},
  \bibinfo {author} {\bibfnamefont {J.~T.}\ \bibnamefont {Daub}}, \bibinfo
  {author} {\bibfnamefont {F.-K.}\ \bibnamefont {Guo}}, \bibinfo {author}
  {\bibfnamefont {C.}~\bibnamefont {Hanhart}}, \bibinfo {author} {\bibfnamefont
  {B.}~\bibnamefont {Kubis}}, \bibinfo {author} {\bibfnamefont {U.-G.}\
  \bibnamefont {Mei\ss{}ner}}, \ and\ \bibinfo {author} {\bibfnamefont {B.-S.}\
  \bibnamefont {Zou}},\ }\href {\doibase 10.1103/PhysRevD.95.034022} {\bibfield
   {journal} {\bibinfo  {journal} {Phys. Rev. D}\ }\textbf {\bibinfo {volume}
  {95}},\ \bibinfo {pages} {034022} (\bibinfo {year} {2017})},\ \Eprint
  {http://arxiv.org/abs/1611.00913} {arXiv:1611.00913 [hep-ph]} \BibitemShut
  {NoStop}%
\bibitem [{\citenamefont {Garcia-Martin}\ \emph {et~al.}(2011)\citenamefont
  {Garcia-Martin}, \citenamefont {Kaminski}, \citenamefont {Pelaez},
  \citenamefont {Ruiz~de Elvira},\ and\ \citenamefont
  {Yndurain}}]{GarciaMartin:2011cn}%
  \BibitemOpen
  \bibfield  {author} {\bibinfo {author} {\bibfnamefont {R.}~\bibnamefont
  {Garcia-Martin}}, \bibinfo {author} {\bibfnamefont {R.}~\bibnamefont
  {Kaminski}}, \bibinfo {author} {\bibfnamefont {J.~R.}\ \bibnamefont
  {Pelaez}}, \bibinfo {author} {\bibfnamefont {J.}~\bibnamefont {Ruiz~de
  Elvira}}, \ and\ \bibinfo {author} {\bibfnamefont {F.~J.}\ \bibnamefont
  {Yndurain}},\ }\href {\doibase 10.1103/PhysRevD.83.074004} {\bibfield
  {journal} {\bibinfo  {journal} {Phys. Rev. D}\ }\textbf {\bibinfo {volume}
  {83}},\ \bibinfo {pages} {074004} (\bibinfo {year} {2011})},\ \Eprint
  {http://arxiv.org/abs/1102.2183} {arXiv:1102.2183 [hep-ph]} \BibitemShut
  {NoStop}%
\bibitem [{\citenamefont {Pelaez}\ and\ \citenamefont
  {Rodas}(2018)}]{Pelaez:2018qny}%
  \BibitemOpen
  \bibfield  {author} {\bibinfo {author} {\bibfnamefont {J.~R.}\ \bibnamefont
  {Pelaez}}\ and\ \bibinfo {author} {\bibfnamefont {A.}~\bibnamefont {Rodas}},\
  }\href {\doibase 10.1140/epjc/s10052-018-6296-9} {\bibfield  {journal}
  {\bibinfo  {journal} {Eur. Phys. J. C}\ }\textbf {\bibinfo {volume} {78}},\
  \bibinfo {pages} {897} (\bibinfo {year} {2018})},\ \Eprint
  {http://arxiv.org/abs/1807.04543} {arXiv:1807.04543 [hep-ph]} \BibitemShut
  {NoStop}%
\bibitem [{\citenamefont {Descotes-Genon}(2000)}]{Descotes-Genon:2000pfd}%
  \BibitemOpen
  \bibfield  {author} {\bibinfo {author} {\bibfnamefont {S.}~\bibnamefont
  {Descotes-Genon}},\ }\emph {\bibinfo {title} {{Effet des boucles de quarks
  l\'egers sur la structure chirale du vide de QCD}}},\ \href@noop {} {Ph.D.
  thesis},\ \bibinfo  {school} {U. Paris-Sud 11, Dept. Phys., Orsay} (\bibinfo
  {year} {2000})\BibitemShut {NoStop}%
\bibitem [{\citenamefont {Peset}\ \emph
  {et~al.}(2018{\natexlab{b}})\citenamefont {Peset}, \citenamefont {Pineda},\
  and\ \citenamefont {Segovia}}]{Peset:2018ria}%
  \BibitemOpen
  \bibfield  {author} {\bibinfo {author} {\bibfnamefont {C.}~\bibnamefont
  {Peset}}, \bibinfo {author} {\bibfnamefont {A.}~\bibnamefont {Pineda}}, \
  and\ \bibinfo {author} {\bibfnamefont {J.}~\bibnamefont {Segovia}},\ }\href
  {\doibase 10.1007/JHEP09(2018)167} {\bibfield  {journal} {\bibinfo  {journal}
  {JHEP}\ }\textbf {\bibinfo {volume} {09}},\ \bibinfo {pages} {167} (\bibinfo
  {year} {2018}{\natexlab{b}})},\ \Eprint {http://arxiv.org/abs/1806.05197}
  {arXiv:1806.05197 [hep-ph]} \BibitemShut {NoStop}%
\bibitem [{\citenamefont {Chetyrkin}\ \emph {et~al.}(2000)\citenamefont
  {Chetyrkin}, \citenamefont {Kuhn},\ and\ \citenamefont
  {Steinhauser}}]{Chetyrkin:2000yt}%
  \BibitemOpen
  \bibfield  {author} {\bibinfo {author} {\bibfnamefont {K.~G.}\ \bibnamefont
  {Chetyrkin}}, \bibinfo {author} {\bibfnamefont {J.~H.}\ \bibnamefont {Kuhn}},
  \ and\ \bibinfo {author} {\bibfnamefont {M.}~\bibnamefont {Steinhauser}},\
  }\href {\doibase 10.1016/S0010-4655(00)00155-7} {\bibfield  {journal}
  {\bibinfo  {journal} {Comput. Phys. Commun.}\ }\textbf {\bibinfo {volume}
  {133}},\ \bibinfo {pages} {43} (\bibinfo {year} {2000})},\ \Eprint
  {http://arxiv.org/abs/hep-ph/0004189} {arXiv:hep-ph/0004189} \BibitemShut
  {NoStop}%
\bibitem [{\citenamefont {Chivukula}\ \emph {et~al.}(1989)\citenamefont
  {Chivukula}, \citenamefont {Cohen}, \citenamefont {Georgi}, \citenamefont
  {Grinstein},\ and\ \citenamefont {Manohar}}]{Chivukula:1989ds}%
  \BibitemOpen
  \bibfield  {author} {\bibinfo {author} {\bibfnamefont {R.~S.}\ \bibnamefont
  {Chivukula}}, \bibinfo {author} {\bibfnamefont {A.~G.}\ \bibnamefont
  {Cohen}}, \bibinfo {author} {\bibfnamefont {H.}~\bibnamefont {Georgi}},
  \bibinfo {author} {\bibfnamefont {B.}~\bibnamefont {Grinstein}}, \ and\
  \bibinfo {author} {\bibfnamefont {A.~V.}\ \bibnamefont {Manohar}},\ }\href
  {\doibase 10.1016/0003-4916(89)90119-X} {\bibfield  {journal} {\bibinfo
  {journal} {Annals Phys.}\ }\textbf {\bibinfo {volume} {192}},\ \bibinfo
  {pages} {93} (\bibinfo {year} {1989})}\BibitemShut {NoStop}%
\bibitem [{\citenamefont {Besson}\ \emph {et~al.}(1985)\citenamefont {Besson}
  \emph {et~al.}}]{Besson:1984bd}%
  \BibitemOpen
  \bibfield  {author} {\bibinfo {author} {\bibfnamefont {D.}~\bibnamefont
  {Besson}} \emph {et~al.} (\bibinfo {collaboration} {CLEO}),\ }\href {\doibase
  10.1103/PhysRevLett.54.381} {\bibfield  {journal} {\bibinfo  {journal} {Phys.
  Rev. Lett.}\ }\textbf {\bibinfo {volume} {54}},\ \bibinfo {pages} {381}
  (\bibinfo {year} {1985})}\BibitemShut {NoStop}%
\bibitem [{\citenamefont {Lovelock}\ \emph {et~al.}(1985)\citenamefont
  {Lovelock} \emph {et~al.}}]{Lovelock:1985nb}%
  \BibitemOpen
  \bibfield  {author} {\bibinfo {author} {\bibfnamefont {D.~M.~J.}\
  \bibnamefont {Lovelock}} \emph {et~al.},\ }\href {\doibase
  10.1103/PhysRevLett.54.377} {\bibfield  {journal} {\bibinfo  {journal} {Phys.
  Rev. Lett.}\ }\textbf {\bibinfo {volume} {54}},\ \bibinfo {pages} {377}
  (\bibinfo {year} {1985})}\BibitemShut {NoStop}%
\bibitem [{\citenamefont {Colangelo}\ \emph {et~al.}(2018)\citenamefont
  {Colangelo}, \citenamefont {Lanz}, \citenamefont {Leutwyler},\ and\
  \citenamefont {Passemar}}]{Colangelo:2018jxw}%
  \BibitemOpen
  \bibfield  {author} {\bibinfo {author} {\bibfnamefont {G.}~\bibnamefont
  {Colangelo}}, \bibinfo {author} {\bibfnamefont {S.}~\bibnamefont {Lanz}},
  \bibinfo {author} {\bibfnamefont {H.}~\bibnamefont {Leutwyler}}, \ and\
  \bibinfo {author} {\bibfnamefont {E.}~\bibnamefont {Passemar}},\ }\href
  {\doibase 10.1140/epjc/s10052-018-6377-9} {\bibfield  {journal} {\bibinfo
  {journal} {Eur. Phys. J. C}\ }\textbf {\bibinfo {volume} {78}},\ \bibinfo
  {pages} {947} (\bibinfo {year} {2018})},\ \Eprint
  {http://arxiv.org/abs/1807.11937} {arXiv:1807.11937 [hep-ph]} \BibitemShut
  {NoStop}%
\bibitem [{\citenamefont {Voloshin}\ and\ \citenamefont
  {Zakharov}(1980)}]{Voloshin:1980zf}%
  \BibitemOpen
  \bibfield  {author} {\bibinfo {author} {\bibfnamefont {M.~B.}\ \bibnamefont
  {Voloshin}}\ and\ \bibinfo {author} {\bibfnamefont {V.~I.}\ \bibnamefont
  {Zakharov}},\ }\href {\doibase 10.1103/PhysRevLett.45.688} {\bibfield
  {journal} {\bibinfo  {journal} {Phys. Rev. Lett.}\ }\textbf {\bibinfo
  {volume} {45}},\ \bibinfo {pages} {688} (\bibinfo {year} {1980})}\BibitemShut
  {NoStop}%
\bibitem [{\citenamefont {Novikov}\ and\ \citenamefont
  {Shifman}(1981)}]{Novikov:1980fa}%
  \BibitemOpen
  \bibfield  {author} {\bibinfo {author} {\bibfnamefont {V.~A.}\ \bibnamefont
  {Novikov}}\ and\ \bibinfo {author} {\bibfnamefont {M.~A.}\ \bibnamefont
  {Shifman}},\ }\href {\doibase 10.1007/BF01429829} {\bibfield  {journal}
  {\bibinfo  {journal} {Z. Phys. C}\ }\textbf {\bibinfo {volume} {8}},\
  \bibinfo {pages} {43} (\bibinfo {year} {1981})}\BibitemShut {NoStop}%
\bibitem [{\citenamefont {Stern}\ \emph {et~al.}(1993)\citenamefont {Stern},
  \citenamefont {Sazdjian},\ and\ \citenamefont {Fuchs}}]{Stern:1993rg}%
  \BibitemOpen
  \bibfield  {author} {\bibinfo {author} {\bibfnamefont {J.}~\bibnamefont
  {Stern}}, \bibinfo {author} {\bibfnamefont {H.}~\bibnamefont {Sazdjian}}, \
  and\ \bibinfo {author} {\bibfnamefont {N.~H.}\ \bibnamefont {Fuchs}},\ }\href
  {\doibase 10.1103/PhysRevD.47.3814} {\bibfield  {journal} {\bibinfo
  {journal} {Phys. Rev. D}\ }\textbf {\bibinfo {volume} {47}},\ \bibinfo
  {pages} {3814} (\bibinfo {year} {1993})},\ \Eprint
  {http://arxiv.org/abs/hep-ph/9301244} {arXiv:hep-ph/9301244} \BibitemShut
  {NoStop}%
\bibitem [{\citenamefont {Watson}(1954)}]{Watson:1954uc}%
  \BibitemOpen
  \bibfield  {author} {\bibinfo {author} {\bibfnamefont {K.~M.}\ \bibnamefont
  {Watson}},\ }\href {\doibase 10.1103/PhysRev.95.228} {\bibfield  {journal}
  {\bibinfo  {journal} {Phys. Rev.}\ }\textbf {\bibinfo {volume} {95}},\
  \bibinfo {pages} {228} (\bibinfo {year} {1954})}\BibitemShut {NoStop}%
\bibitem [{\citenamefont {Muskhelishvili}(1958)}]{mushi}%
  \BibitemOpen
  \bibfield  {author} {\bibinfo {author} {\bibfnamefont {N.}~\bibnamefont
  {Muskhelishvili}},\ }\href@noop {} {\emph {\bibinfo {title} {Singular
  integral equations}}}\ (\bibinfo  {publisher} {Noordhoff},\ \bibinfo
  {address} {Groningen},\ \bibinfo {year} {1958})\BibitemShut {NoStop}%
\bibitem [{\citenamefont {Omnes}(1958)}]{Omnes:1958hv}%
  \BibitemOpen
  \bibfield  {author} {\bibinfo {author} {\bibfnamefont {R.}~\bibnamefont
  {Omnes}},\ }\href {\doibase 10.1007/BF02747746} {\bibfield  {journal}
  {\bibinfo  {journal} {Nuovo Cim.}\ }\textbf {\bibinfo {volume} {8}},\
  \bibinfo {pages} {316} (\bibinfo {year} {1958})}\BibitemShut {NoStop}%
\bibitem [{\citenamefont {Chen}\ \emph {et~al.}(2016)\citenamefont {Chen},
  \citenamefont {Daub}, \citenamefont {Guo}, \citenamefont {Kubis},
  \citenamefont {Mei\ss{}ner},\ and\ \citenamefont {Zou}}]{Chen:2015jgl}%
  \BibitemOpen
  \bibfield  {author} {\bibinfo {author} {\bibfnamefont {Y.-H.}\ \bibnamefont
  {Chen}}, \bibinfo {author} {\bibfnamefont {J.~T.}\ \bibnamefont {Daub}},
  \bibinfo {author} {\bibfnamefont {F.-K.}\ \bibnamefont {Guo}}, \bibinfo
  {author} {\bibfnamefont {B.}~\bibnamefont {Kubis}}, \bibinfo {author}
  {\bibfnamefont {U.-G.}\ \bibnamefont {Mei\ss{}ner}}, \ and\ \bibinfo {author}
  {\bibfnamefont {B.-S.}\ \bibnamefont {Zou}},\ }\href {\doibase
  10.1103/PhysRevD.93.034030} {\bibfield  {journal} {\bibinfo  {journal} {Phys.
  Rev. D}\ }\textbf {\bibinfo {volume} {93}},\ \bibinfo {pages} {034030}
  (\bibinfo {year} {2016})},\ \Eprint {http://arxiv.org/abs/1512.03583}
  {arXiv:1512.03583 [hep-ph]} \BibitemShut {NoStop}%
\end{thebibliography}%

\end{document}